\documentclass{aa}
\usepackage{amsmath}
\usepackage{graphicx,color}
\usepackage{natbib}
\usepackage{microtype}
\usepackage{url}
\usepackage[varg]{txfonts}
\usepackage{threeparttable}
\usepackage{booktabs, makecell}

\newcommand{\be}{\begin{equation}}
\newcommand{\ee}{\end{equation}}
\newcommand{\bea}{\begin{eqnarray}}
\newcommand{\eea}{\end{eqnarray}}

\newcommand{\nh}{N_{\rm H}}

\def\asca{\textit{ASCA}\xspace}
\def\cxo{\textit{Chandra}\xspace}
\def\ero{\textit{eROSITA}\xspace}
\def\ede{{eROSITA\_DE}\xspace}

\def\nic{\textit{NICER}\xspace}
\def\nus{\textit{NuSTAR}\xspace}
\def\ros{\textit{ROSAT}\xspace}
\def\srgero{\textit{SRG}/eROSITA\xspace}
\def\srg{\textit{SRG}\xspace}

\def\xmmn{\textit{XMM-Newton}\xspace}
\def\xspec{\texttt{\textit{Xspec}}\xspace}

\newcommand\psr{{PSR\,B0656$+$14}\xspace}

\begin{document}

\title{Phase-resolved X-ray spectroscopy of \psr\ with \srgero\ and \xmmn\thanks{Based on observations obtained with \xmmn, an ESA science mission with instruments and contributions directly funded by ESA Member States and NASA}}

\author{Axel Schwope\inst{1}
\and Adriana M. Pires \inst{1,2}
\and Jan Kurpas\inst{1,3}
\and Victor Doroshenko\inst{4}
\and Valery F. Suleimanov\inst{4,5,6}
\and Michael Freyberg\inst{7}
\and Werner Becker\inst{7}
\and Konrad Dennerl\inst{7}
\and Frank Haberl\inst{7}
\and Georg Lamer\inst{1}
\and Chandreyee Maitra\inst{7}
\and Alexander Y. Potekhin\inst{8}
\and Miriam E.~Ramos-Ceja\inst{7}
\and Andrea Santangelo\inst{4}
\and Iris Traulsen\inst{1}
\and Klaus Werner\inst{4}
}
\institute{Leibniz-Institut f\"ur Astrophysik Potsdam (AIP), An der Sternwarte 16, 14482 Potsdam, Germany\\
\email{aschwope@aip.de}
\and
   Purple Mountain Observatory, Key Laboratory of Dark Matter and Space Astronomy, Chinese Academy of Sciences, Nanjing 210008, China
\and 
Potsdam University, Institute for Physics and Astronomy, Karl-Liebknecht-Straße 24/25, 14476 Potsdam, Germany
\and
Institut f\"ur Astronomie und Astrophysik, Sand 1, 72076 Tübingen, Germany
\and 
Astronomy Department, Kazan (Volga region) Federal University, Kremlyovskaya str. 18, 420008 Kazan, Russia
\and 
Space Research Institute of the Russian Academy of Sciences, Profsoyuznaya str. 84/32, 117997 Moscow, Russia 
\and
     Max-Planck-Institut f\"ur extraterrestrische Physik,
     Gie{\ss}enbachstra{\ss}e 1, 85748 Garching, Germany
\and
Ioffe Institute, Politekhnicheskaya 26, 194021, Saint Petersburg, Russia
}
\authorrunning{Schwope et al.}
\titlerunning{Spin-phase resolved spectroscopy of \psr with \srgero}
\date{\today}


\abstract
{We present a detailed spectroscopic and timing analysis of X-ray observations of the bright radio-to-gamma-ray emitting pulsar \psr, which were obtained simultaneously with \ero and \xmmn during the Calibration and Performance Verification phase of the Spektrum-Roentgen-Gamma mission (SRG). The analysis of the 100\,ks-deep observation of \ero is supported by archival and simultaneous observations of the source, including \xmmn, \nus, and \nic. Using \xmmn and \nic we firstly established an X-ray ephemeris for the time interval 2015 to 2020, which connects all X-ray observations in this period without cycle count alias and phase shifts. The mean \ero spectrum clearly reveals an absorption feature originating from the star at 570\,eV with a Gaussian $\sigma$ of about 70\,eV, tentatively identified earlier in a long \xmmn observation \citep{arumugasamy+18}. A second absorption feature, described here as an absorption edge, occurs at $260-265$\,eV. It could be of atmospheric or of instrumental origin. These absorption features are superposed on various emission components, phenomenologically described here as the sum of hot (120\,eV) and cold (65\,eV) blackbody components, both of photospheric origin, and a power-law with photon index $\Gamma=2$ from the magnetosphere. We created energy-dependent lightcurves and phase-resolved spectra with high signal-to-noise. The phase-resolved spectroscopy reveals that the Gaussian absorption line at 570\,eV is clearly present throughout $\sim$60\% of the spin cycle, but undetected otherwise. Likewise, its parameters were found to be dependent on phase. The visibility of the line strength coincides in phase with the maximum flux of the hot blackbody. If the line originates from the stellar surface, it nevertheless likely originates from a different location than the hot polar cap. We also present three families of model atmospheres: a magnetised atmosphere, a condensed surface, and a mixed model, which were applied to the mean observed spectrum and whose continuum fit the observed data well. The atmosphere model, however, predicts too short distances. For the mixed model, the Gaussian absorption may be interpreted as proton cyclotron absorption in a field as high as $10^{14}$\,G, which is significantly higher than that derived from the moderate observed spin-down.
}

\maketitle

\section{Introduction\label{s:intro}}

\psr\ (hereafter B0656) is a middle-aged pulsar that was observed at all wavelengths from the radio regime up to gamma energies. It is the brightest of the ``Three Musketeers'', nicknamed as such by \cite{becker+truemper97} due to their exceptional brightness and similar periods, ages, and spectral energy distribution (SED). It was one of the 27 rotation-powered pulsars detected already with \ros \citep{becker+truemper97} and shows thermal and non-thermal radiation components. 

Initially the thermal component was modeled as a blackbody but already \cite{possenti+96} found the \ros\ spectrum to be more complex. The combined \ros/\asca X-ray spectrum then revealed a phenomenological model valid with modifications till today that consists of two blackbody components from the stellar surface and a power law tail of magnetospheric origin \citep{greiveldinger+96}. This model was confirmed by CCD and grating spectroscopic observations performed with \cxo\ by \citet{pavlov+02}. Using the \cxo\ grating (LETG) spectrum, \cite{marshall+schulz02} stressed that no absorption line was found in the spectral range $0.2-1.0$\,keV. Their phase-average spectrum could be well represented with the sum of two blackbodies. First \xmmn\ observations, both in timing and imaging (small window) mode, were presented by \cite{deluca+05}, confirming the earlier spectral model. The much improved photon counting statistics allowed for the first time to trace the various spectral components  through the spin cycle of the $P_{\rm rot} \simeq 385$\,msec pulsar with the two blackbody-components varying in antiphase. Recently, \cite{arumugasamy+18} reported a possible phase-dependent absorption feature with central energy between $0.5-0.6$\,keV in coordinated \xmmn/\nus-observations. The inclusion of a Gaussian absorption line for roughly 50\% of the spin cycle improved the phenomenological model.

\cite{harding+19} report a summary of \nic observations showing three distinct hot spots that cover different energy bands and rotational phases: a cool thermal radiation component from the entire neutron star surface, a smaller hot spot presumably from polar cap heating, and an additional one at intermediate temperature. The X-ray emission peaks from these hot spots occur at different rotation phases, that are also different from the phases of the radio and gamma-ray peaks. The complex variation of temperature across the surface possibly suggests evolution of multipolar magnetic field structure.

\begin{figure*}[t]
\resizebox{\hsize}{!}{\includegraphics{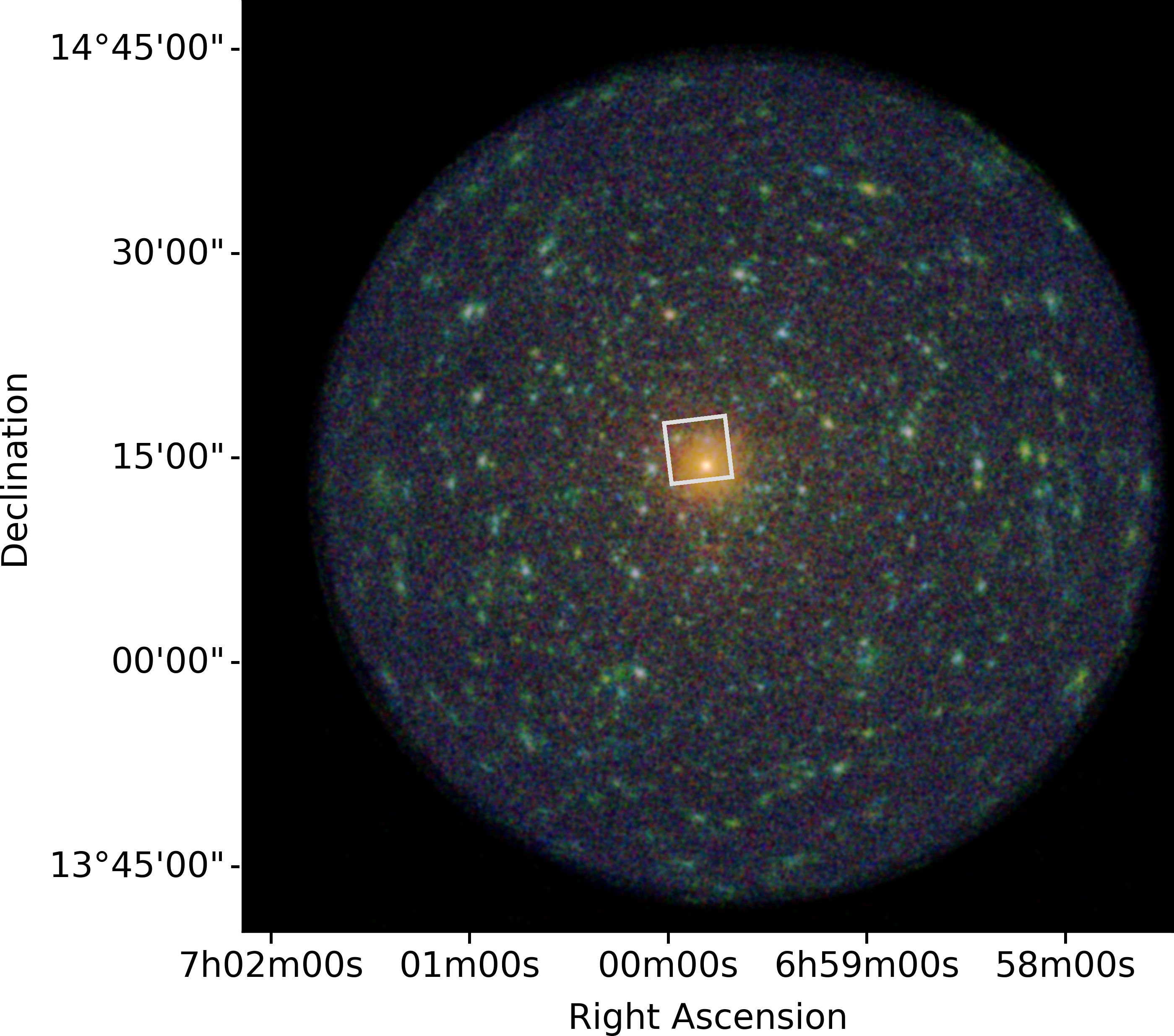}}
\caption{Composite \ero image produced from selected photons in three different energy bands (red: 0.2 - 0.5 keV, green: 0.5 - 1.5 keV, blue: 1.5 - 7 keV) showing the large \ero field of view. The pulsar B0656 is centrally located at the aimpoint. The white rectangle shows the field of view of the \xmmn\ observations that were conducted with EPIC-pn in the Small Window Mode simultaneously.
\label{f:fov}
}
\end{figure*}

\begin{table*}[t]
\caption{Joint and supporting \srgero, \xmmn, and \nus observations of \psr
\label{t:obslog}}
\centering
\begin{tabular}{lrcccccr}
\hline\hline
Observatory & \multicolumn{1}{c}{ObsID} & Instrument & \multicolumn{2}{c}{Start Time} & \multicolumn{2}{c}{End Time} & Exposure\tablefootmark{a} \\
 & & & \multicolumn{2}{c}{(UTC)} & \multicolumn{2}{c}{(UTC)} & \multicolumn{1}{c}{(ks)} \\
\hline
\srgero & 300000 & TM234567 & 2019-Oct-14 & 08:59:40 & 2019-Oct-15 & 17:00:10 & 98.5 \\
\xmmn & 0853000201 & EPIC pn, RGS & 2019-Oct-14 & 12:57:07 & 2019-Oct-15 & 09:01:07 & 71.0 \\
\xmmn & 0762890101 & EPIC pn, RGS & 2015-Sep-19 & 20:18:45 & 2019-Oct-21 & 08:00:03 & 87.1 \\
\xmmn & 0112200101 & EPIC pn & 2001-Oct-23 & 13:45:23 & 2001-Oct-23 & 20:48:24 & 24.4 \\
\nus & 40101004002  & FPMA, FPMB & 2015-Sep-19 & 00:16:08 & 2015-Sep-22 & 08:21:08 & 126.8 \\
\hline
\end{tabular}
\tablefoot{Additional supporting \nic observations of the target are documented in detail in Table~\ref{t:xraytoa}. 
\tablefoottext{a}{Net exposure in kiloseconds excluding time intervals of high background activity and instrument overheads, averaged over active instrument.}}
\end{table*}

The most recent account of the phase-averaged X-ray spectrum of B0656 was given by \cite{zharikov+21}, who on the one hand confirmed the earlier spectral model by \cite{arumugasamy+18}, but also refined it through the inclusion of infrared/optical/ultraviolet data (IR/opt/UV). The spectral energy distribution established in this paper is best described by a broken power law from the magnetosphere plus the double blackbody with superposed absorption line at $\sim$0.5\,keV originating from the stellar surface. The inclusion of the low-energy data hints to a smaller NS radius than obtained from just the X-ray spectral fits, which otherwise overpredict the observed IR/opt/UV emission from the object.

The possible existence of an absorption line in the soft part of the spectrum triggered the \srgero observation of B0656 in its {\it Calibration and Performance Verification Phase} for 100\,ks. At that early phase in the SRG mission, calibration uncertainties were to be expected. We thus asked the \xmmn project scientist to support the SRG mission and this observation in particular with a simultaneous observation of our target in Director's Discretionary Time. Thankfully, this time was granted and the two observatories obtained data simultaneously. 

These recent \xmmn data were presented already by \cite{zharikov+21} for their analysis of the mean spectrum of the source. They are put in context with \ero in the current paper, where a spin-phase resolved analysis is presented.
Here we focus on the spectral analysis of the coordinated \ero/\xmmn-observations, but take other X-ray data into account to describe the overall spectral energy distribution (including \nus) and to improve the timing solution {\nic} of the pulsar. We also briefly address calibration items as far as the timing system is concerned. 

The targeted observations of B0656, some 8\,deg above the galactic plane, give at the same time one of the deepest looks into the Galaxy by \ero\ in the years to come. Given the 63\arcmin\ field of view, such a deep observation is a full fledged X-ray survey on its own right. Indeed, more than 960 X-ray sources were found serendipitously in the same observation ($0.6-2.3$\,keV; all telescope modules). The comprehensive source catalogue from this observation is published in a separate paper \citep{lamer+21}.

The paper is structured as follows. In Sect.~\ref{s:obs} we describe the analysed observational data set and the data reduction. The results are detailed in Sect.~\ref{s:results}. In Sect.~\ref{s:timing} we present a new timing solution for the pulsar and analysis of the lightcurves. 
A spectral analysis of both the phase-averaged and the phase-resolved spectra using phenomenological and more physically motivated neutron star atmosphere and condensed surface models is in Sect.~\ref{s:spec}. Our main results are summarised and discussed in Sect.~\ref{s:disc}. We provide additional timing and spectral information on the \ero data in the Appendix.
When computing dimensions of emitting areas and luminosity we use the distance to B0656 from the VLA radio parallax of $288^{+33}_{-27}$\,pc \citep{brisken+03}.

\section{Observations and reductions\label{s:obs}}

\subsection{\srgero\label{s:erosreduction}}

The observations of \psr with \srgero were the first PV-phase observations to be conducted with all seven telescope modules (TM) of the German \ero collaboration eROSITA\_DE after an extended commissioning and calibration phase of the instrument and the observatory as whole. They were performed on October 14, 2019 for a nominal exposure of 100\,ks (see Table \ref{t:obslog}). Two anomalies occurred: TM1 did not reveal science-grade data and a time shift of the central spacecraft clock of $-0.345$\,s was applied during the observations of our target at the onboard execution time of 2019-10-14\ 23:00:45 (UT; cf.~Sect.~\ref{s:tm1} and \ref{s:timeshift}, for more information).

All \ero cameras were operated in FrameStore mode with PMWORK and FILTER setup, the data were reduced and analysed with the eSASS software system eSASSusers\_201009\footnote{For a description of the eSASS tasks and algorithms, see \cite{brunner+21}} within the $0.2-10$\,keV energy band. 
TMs 1, 2, 3, 4, and 6 are operated with an aluminium filter directly deposited on the CCD while TM5 and TM7 are operated with a filter that consists of a polyimide foil with an aluminium layer. The thickness of the aluminium layer is lower, which gives TM5 and TM7 higher sensitivity, in particular at soft X-ray energies. 
Interestingly, these two also suffer from time-variable light leaks which (may) impact their usability for soft X-ray studies \citep[see][for further details]{predehl+21}. 

In the following, the summed signal from the cameras with on-chip filter is referred to as (virtual) telescope module TM8 (here without TM1), the summed signal from the cameras without the filters is referred to as TM9, and the sum of all physical TMs as TM0 (again, in this paper without TM1). At the time of writing, the energy calibration of TM9 is less reliable than that of TM8, disregarding if a particular observation is affected by light leaks. The final spectral results presented below are therefore derived for TM8 in the main body of the paper. For cross-calibration purposes, the results of individual TMs and TM0 and TM9 are reported in Appendix~\ref{s:eroccalib}. Our timing analysis is based on all active detectors.

\begin{figure*}[t]
\resizebox{0.49\hsize}{!}{\includegraphics{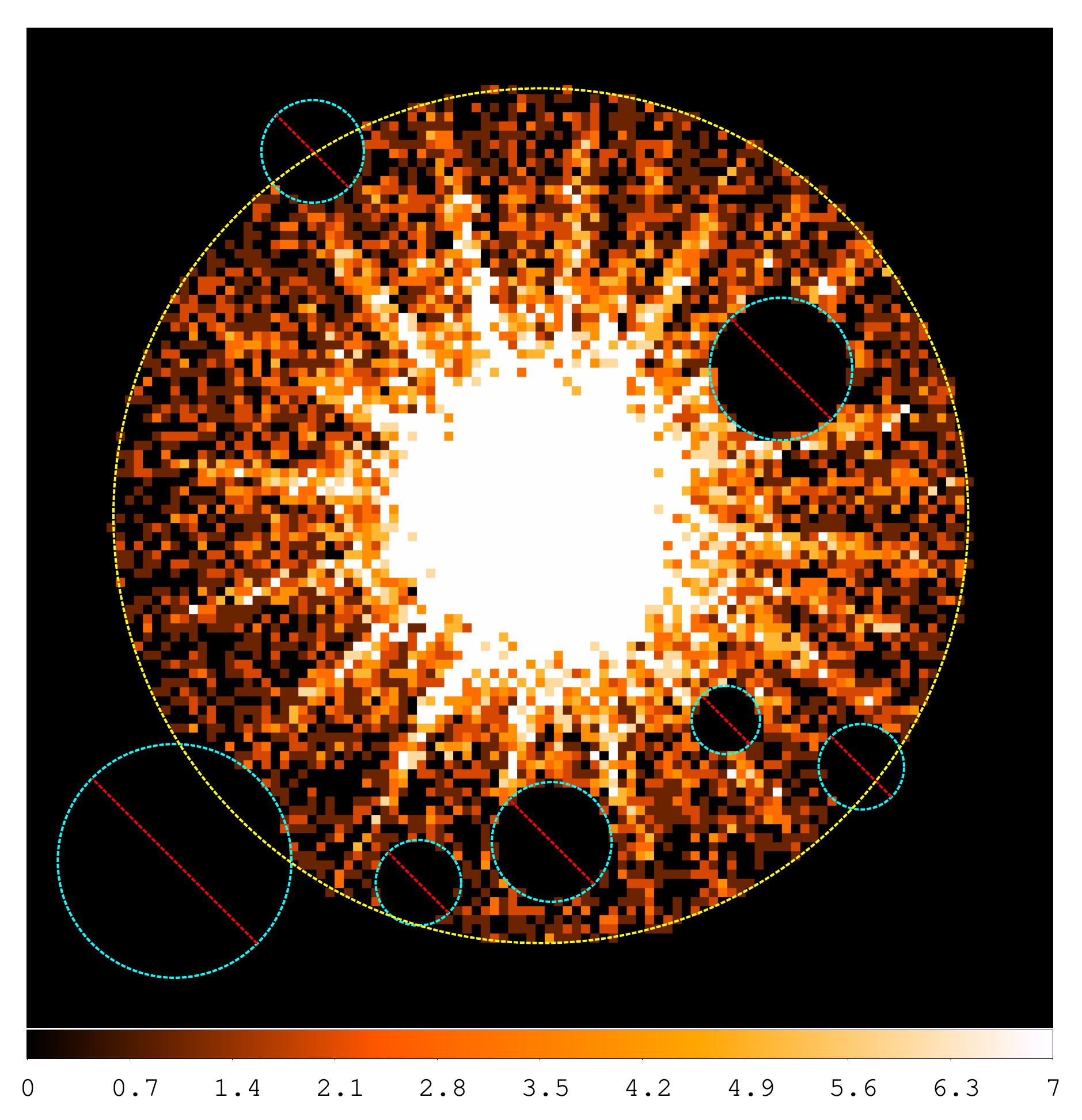}}
\resizebox{0.49\hsize}{!}{\includegraphics{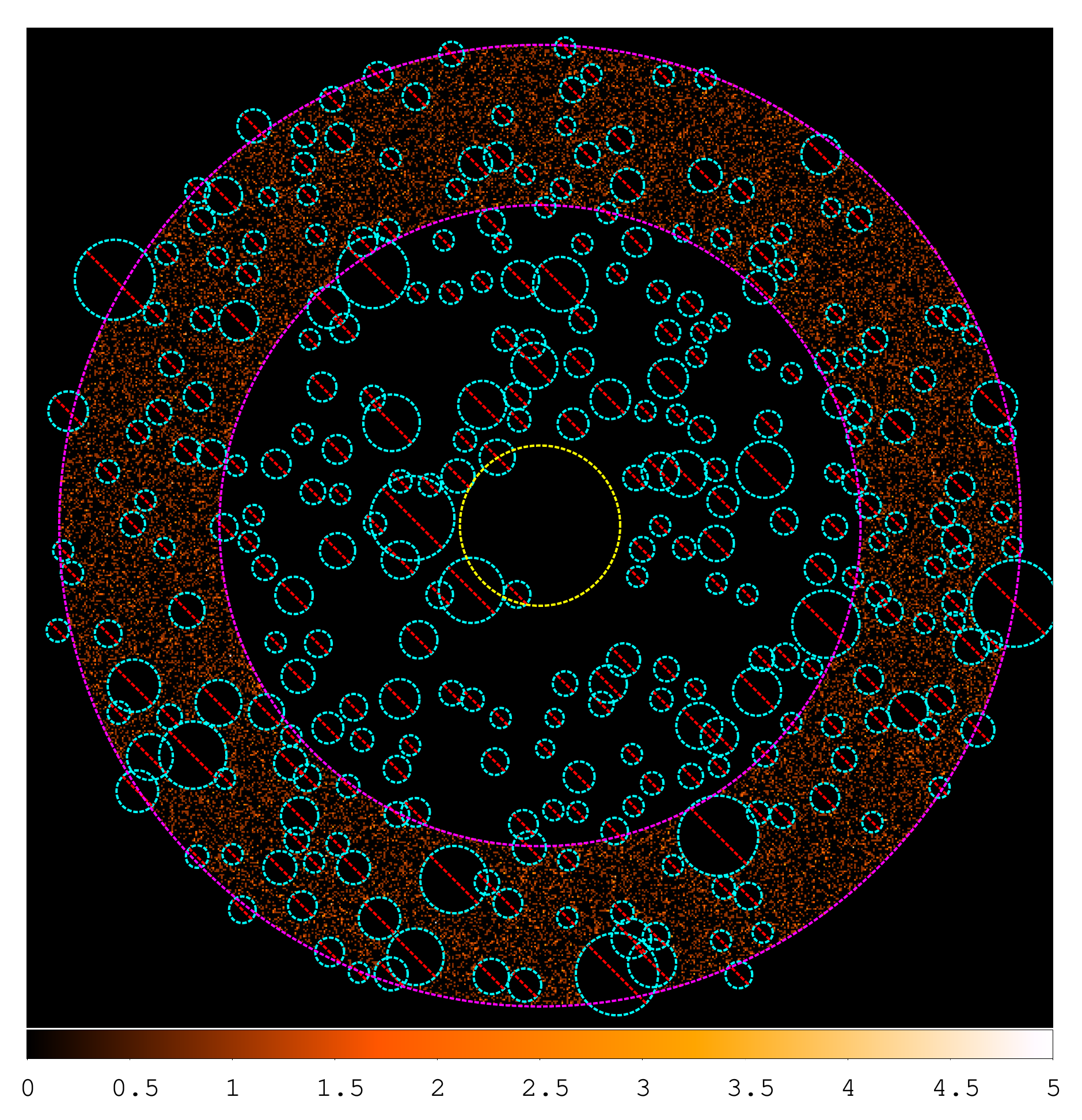}}
\caption{Extraction regions adopted in the analysis. The source and background regions are centred at right ascension 104.951117\,deg and declination +14.239175\,deg. The exclusion zones of contaminants are shown as cyan circles with a red bar. {\it Left}. The source extraction radius (yellow) is 150\arcsec\ (TM2; $0.2-10$\,keV). {\it Right.} The background annulus (magenta) has inner and outer radii 10\arcmin\ and 15\arcmin, respectively (TM6; $0.2-10$\,keV).
\label{f:regi}}
\end{figure*}

The event files were filtered for periods of high background activity with the eSASS task {\texttt{flaregti}. The flaring count rate level (threshold) of each detector for the duration of the observation was determined on the basis of the mean surface brightness of local pixels in a pre-defined spatial grid. We adopted the $2.2-10$\,keV energy band to minimise contamination from the central source and a time bin size of 100\,s; bright sources in the field-of-view were masked by default. The analysis shows typical threshold values per detector around 2\,s$^{-1}$\,deg$^{-2}$, which were then adopted for GTI filtering. Averaged over all telescope modules, the percentage of time loss due to flares is low, about $1.6$\%. 
We inspected the GTI-filtered images, created for all TMs in different energy ranges, for artifacts and possible light leaks. We found that only TM7 shows a pronounced light leak visible at soft energies, while TM5 is seemingly unaffected. 

\begin{table}[t]
\caption{Results of source detection and ML PSF fitting
\label{t:srcdet}}
\centering
\begin{tabular}{l r}
\hline\hline
Parameter &  \\
\hline
\multicolumn{2}{l}{\# X-ray sources (FoV)}\\
\ldots$0.2-0.6$\,keV & 508\\
\ldots$0.6-2.3$\,keV & 968\\
\ldots$2.3-5.0$\,keV & 279\\
Detection likelihood & $5\times10^6$ \\
ML Counts & \\
\ldots$0.2-0.5$\,keV & $5.7258(9)\times10^5$\\
\ldots$0.5-1.0$\,keV & $2.220(3)\times10^5$\\
\ldots$1.0-2.2$\,keV & $1.690(13)\times10^4$\\
\ldots$2.2-10$\,keV  & $4.41(25)\times10^2$\\
Rate (s$^{-1}$) & $7.697(9)$\\
\tablefootmark{$\dagger$}HR$_1$ & $-0.4412(6)$ \\
\tablefootmark{$\dagger$}HR$_2$ & $-0.8585(10)$\\
\tablefootmark{$\dagger$}HR$_3$ & $-0.9492(28)$\\
RA (h min sec) & 6:59:48.3\\
Dec (d m s) &   14:14:20.993\\
\hline
\end{tabular}
\tablefoot{Detection likelihood, count rate, and equatorial coordinates as computed with the task {\it ermldet} in the total \ero energy band, considering all active telescope modules (except for TM1), and all valid patterns. 
\tablefoottext{$\dagger$}{Hardness ratios (HR) are ratios of the difference to total counts in two contiguous of four \ero energy bands: $0.2-0.5$\,keV, $0.5-1$\,keV, $1-2.2$\,keV, and $2.2-10$\,keV.}}
\end{table}

We performed source detection based on maximum likelihood (ML) PSF fitting to generate catalogues in various energy bands \citep[see e.g.][for details]{lamer+21,liu+21}.
For maximal sensitivity, all GTI-filtered event tables of the six telescope modules and all valid photon patterns were considered. In Table~\ref{t:srcdet} we list the number of sources detected in the field-of-view in three broad energy bands (soft, $0.2-0.6$\,keV, medium, $0.6-2.3$\,keV, and hard, $2.3-5$\,keV) and the source characterisation parameters from ML PSF fitting.

The catalogues were then used to optimise the coordinates and sizes of the source and background extraction regions with the ``auto'' option of the {\it srctool} task.
In this mode, the task can also be used to identify neighbouring sources (``contaminants''), whose PSF overlaps the regions of interest.
We chose both the source (circular) and background (annular) regions to be centred around the target coordinates as determined with {\it ermldet} in the soft energy band. Additionally, we screened the catalogues for spurious extended detections, erroneously identified as contaminants in the wings of the target PSF\footnote{Specifically, we removed from the source list objects with extent parameter larger than 70\arcsec, which were detected up to 5\arcmin\ away from the pulsar's position.}. The various energy bands were used to assess the likelihood of the X-ray sources detected within 2.5\arcmin\ away from the aimpoint.

According to this analysis, we adopted for the target an extraction radius of 150\arcsec\ and determined ``exclusion zones'' around the position of seven contaminants detected in the medium and hard energy bands (Fig.~\ref{f:regi}; left). The seven sources have detection likelihood within $12-400$ and counts within $60-3500$ ($0.6-2.3$\,keV). 
Due to the brightness of the pulsar below 2.3\,keV, the blind application of the {\it srctool} functionality leads to a background annulus with inner and outer radii of $\sim4-5.5\arcmin$ and $\sim27-35\arcmin$, respectively. To minimise further (below 2\%) the contamination from the central source, and avoid uncertainties due to vignetting effects at the very edge of the \ero field-of-view, we decided instead to adopt a narrower, 5\arcmin\ wide, background annulus with inner radius of 10\arcmin; all 152 contaminants detected in the $0.6-2.3$\,keV energy band were then excluded from this region (Fig.~\ref{f:regi}; right).

In Fig.~\ref{f:pat} we show the distribution of photon patterns (that is, the relative fraction of singles, doubles, triples and quadruples) as a function of energy for the events detected at aimpoint. 
Only the detectors with on-chip filter are considered. Solid curves shows the expected distribution according to the model in the calibration database. 
The overall trend of the pattern distribution follows the calibrated one and is also consistent with that of other X-ray bright and nearby isolated neutron stars observed during the CalPV phase \citep{pires+21}. 
On the other hand, the lack of singles and excess of doubles and triples with respect to the calibrated distribution of patterns is remarkable within $0.2-2$\,keV. This mismatch is understood to be due to several factors, among them the exact sub-pixel location of the photons in the detector, contamination from detector noise below 0.4\,keV, and the low-energy threshold applied to the data for telemetry reasons (especially important for the detectors affected by the optical leak).

\begin{figure}[t]
\includegraphics[width=0.49\textwidth]{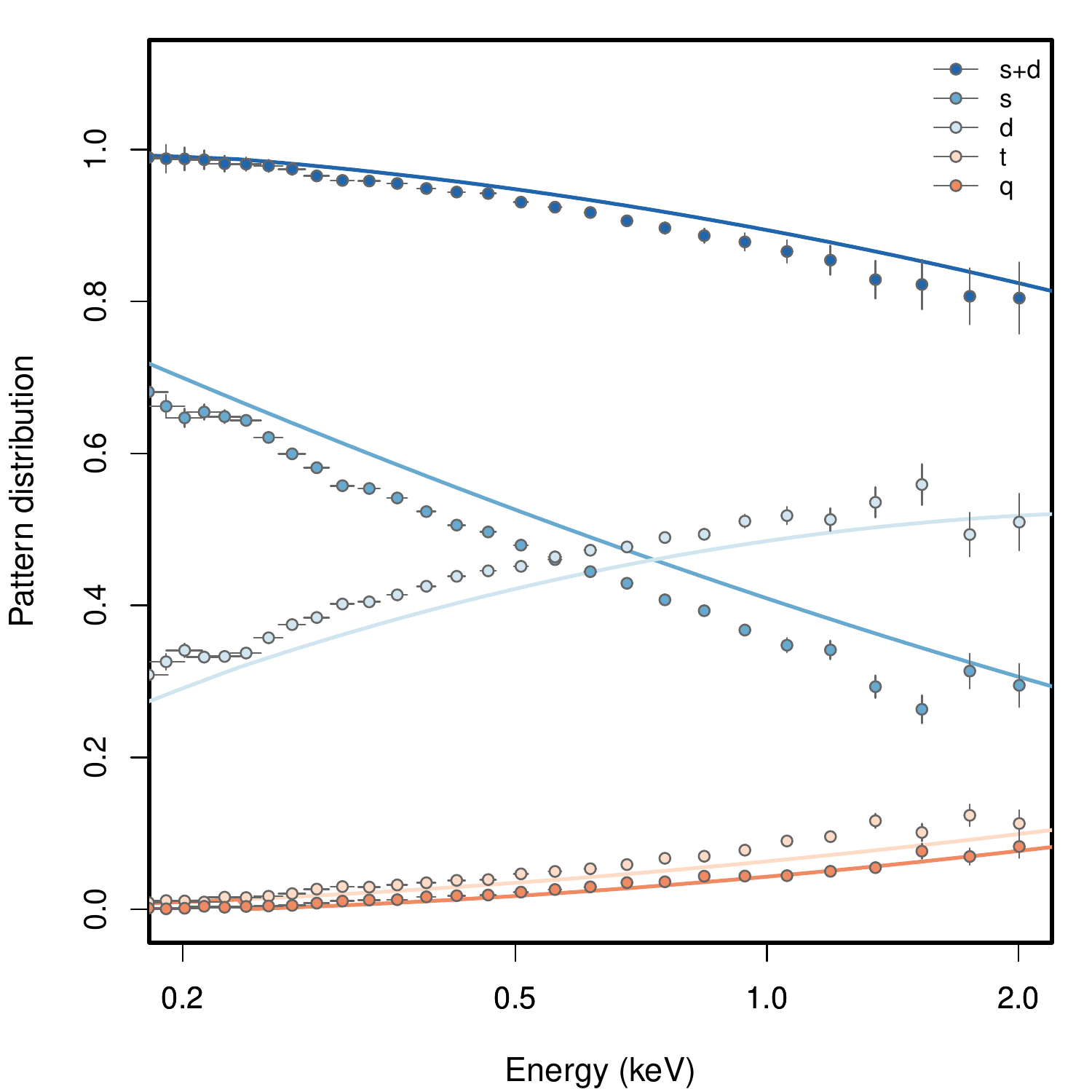}
\caption{Distribution of photon pattern fractions as a function of energy. The observed fractions of singles (s), doubles (d), singles and doubles (s+d), triples (t), and quadruples (q) are shown as data points for the TM8 detector combination. Solid lines show the expected (theoretical) pattern distribution in the $0.2-2$\,keV energy band.\label{f:pat}}
\end{figure}

We extracted lightcurves to measure the pulsar's mean count rate in the $0.2-0.6$\,keV, $0.6-5$\,keV and $0.2-5$\,keV energy bands, using a time bin size of 100\,s. To this end, we used  the flare-filtered event lists and corrected the number of photons in each time bin for the difference in source and background area size, before subtracting the background, while also multiplying them with a psf correction factor computed from the \ero 2dpsf files. 
The resulting mean count rates are shown in Table \ref{t:mcr}. The detectors 5 and 7, which are more sensitive at soft X-rays, show a larger count rate at these energies than TM2, 3, 4 and 6. 
All exposures are consistent with a constant flux.

\begin{table}
\caption{Mean count rates of \psr per \ero detector 
\label{t:mcr}}
\centering
\begin{tabular}{c r r r}
\hline
\hline
Instrument & \multicolumn{3}{c}{Count rate (s$^{-1}$)}\\
\cline{2-4}
 & \tablefootmark{a} & \tablefootmark{b} & \tablefootmark{c} \\
\hline
TM2 & 0.918(3) & 0.2323(16) & 1.151(4)\\
TM3 & 0.934(4) & 0.2338(16) & 1.168(4)\\
TM4 & 0.948(4) & 0.2296(16) & 1.178(4)\\
TM5 & 1.398(4) & 0.2128(16) & 1.610(5)\\
TM6 & 0.920(4) & 0.2311(17) & 1.151(4)\\
TM7 & 1.431(4) & 0.1970(15) & 1.628(5)\\
TM8 & 3.689(8) & 0.919(4) & 4.607(9)\\
TM9 & 2.801(8) & 0.4056(22) & 3.207(8)\\
TM0 & 6.482(12) & 1.323(5) & 7.805(13)\\
\hline
\end{tabular}
\tablefoot{The count rates are measured in the following energy bands:
\tablefoottext{a}{$0.2-0.6$\,keV},
\tablefoottext{b}{$0.6-5$\,keV},
\tablefoottext{c}{$0.2-5$\,keV}.
}
\end{table}

The times-of-arrival of the photons were converted from the local satellite to the solar system barycentric frame using the HEASOFT task {\it barycen}, the JPL-DE405 ephemeris table, the target coordinates from \citet{arumugasamy+18}, and a suitable orbit file covering the epoch of the observation. For the latter, we converted the file with information on the spacecraft position and velocity as provided by NPOL to the FITS format required by {\it barycen}. Likewise, some header keywords in the \ero event files were edited to comply with the tool\footnote{Specifically, we added ``LOCAL'' and ``ICRS'' to the TIMEREF and RADECSYS keywords.}. The correction was performed detector-wise to take into account (and correct) the time stamps and GTI{\it n} extensions of each instrument individually. The corrected event files were then merged back into a master table with separate GTI{\it n} extensions per detector, as is standard in eSASS analysis.

The cleaned and barycentric corrected event lists were used to extract the scientific products -- spectra and phase-folded lightcurves -- adopted further in the analysis (Sect.~\ref{s:results}).
To generate phase-resolved spectra, we split the source photons according to the pulsar's spin phase into multiple event files (see Sect.~\ref{s:pfold}; for details). We kept the original extensions of the pipeline processed event table to ensure compatibility with eSASS; the GTI{\it n} extensions were updated accordingly to cover the respective phases of interest. Finally, the eSASS task {\it srctool} was applied to generate the individual spectrum of each phase interval.

\begin{figure}[t]
\resizebox{\hsize}{!}{\includegraphics{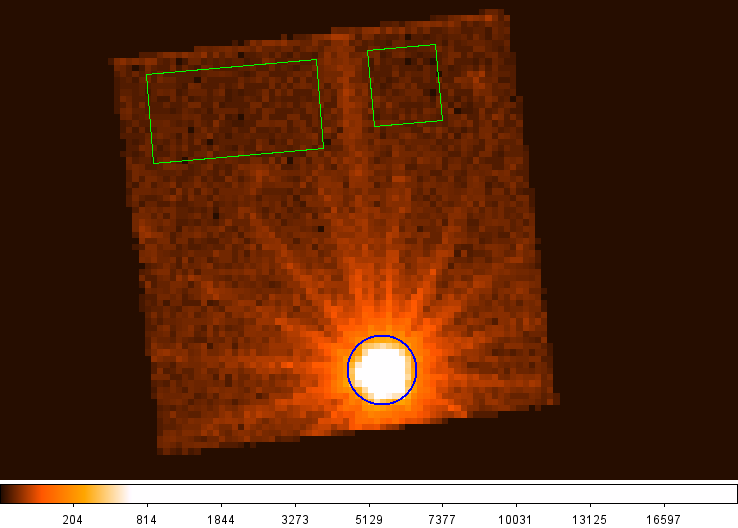}}
\caption{\xmmn SW observation of \psr, performed simultaneously with \ero in October 2019, with overlaid source and background extraction regions (blue circle with radius $22.5\arcsec$ and green boxes of sizes $110\arcsec\times 58\arcsec$ (left) and $44\arcsec\times 50\arcsec$ (right), respectively; 0.3-12\,keV). The image has $63\times 64$ pixels, about $4.3 \times 4.3\arcmin$.
\label{f:xmm}}
\end{figure}

\subsection{\xmmn\label{s:xmmreduction}}

\xmmn observed \psr\ in three occasions in 2001, 2015, and in 2019, the latter simultaneously with \ero \citep[Sect.~\ref{s:intro}; see also][]{deluca+05, arumugasamy+18, zharikov+21}. For the 2019 observation, we chose to observe the pulsar adopting the same instrumental setup as in 2015, that is: the EPIC-pn and MOS instruments in small window (SW) and timing (TI) mode, respectively; all EPIC exposures were performed adopting the THIN1 filter. The RGS1 and RGS2 instruments were used in SES spectroscopy mode. We do not discuss the OM exposures in the analysis presented here.

We reduced the observations with the \xmmn Science Analysis (SAS) software, version 18.0.0, following standard procedure and applying up-to-date calibration files. We extracted event lists for the EPIC instruments using the meta-tasks {\it emproc} and {\it epproc}; the time stamps of the photons, GTI extensions, and time-related header keywords were barycentred with the SAS task {\it barycen}. At both 2015 and 2019 observations, MOS1 did not deliver science-grade data; we do not include MOS2 in the analysis due to its much lower quality with respect to pn as verified by preliminary analysis. 

We adopted a circular source extraction region of radius 22.5\arcsec\ for both pn SW observations. Background events were extracted from two box-like regions on the same CCD as the target, defined so as to avoid out-of-time events along the read-out direction (Fig.~\ref{f:xmm}). We screened the background lightcurves in the energy range $0.3-12$\,keV for flares, adopting a bin size of 50\,s. Following the prescription of \citet{deluca_molendi04}, we defined a ``rejection threshold'' $3\sigma$ from the Gaussian mean, based on the observed distribution of count rates. We used the {\texttt{tabgtigen} task to generate GTI files and used them to filter the event files with {\it evselect}. Only unflagged, single and double events (FLAG = 0 PATTERN $\leq 4$) were selected. The resulting filtered event list was subsequently adopted for the timing and spectral analysis.

For RGS, we used the EPIC source coordinates as determined with the SAS task {\texttt{emldetect}} in each observation to generate the instrument spatial masks and energy filters with the SAS routine {\texttt{rgsproc}}. We identified times of low background activity from the count rate on CCD 9, the closest to the optical axis, and applied a count rate threshold of $0.1$\,s$^{-1}$ to filter the GTIs. Due to electronic problems, one CCD chip of each of the RGS detectors failed early in the mission; these affect the spectral coverage between 11\,\AA\ and 14\,\AA\ and between 20\,\AA\ and 24\,\AA, respectively, in RGS1 and RGS2.

\begin{figure}[t]
\resizebox{\hsize}{!}{\includegraphics{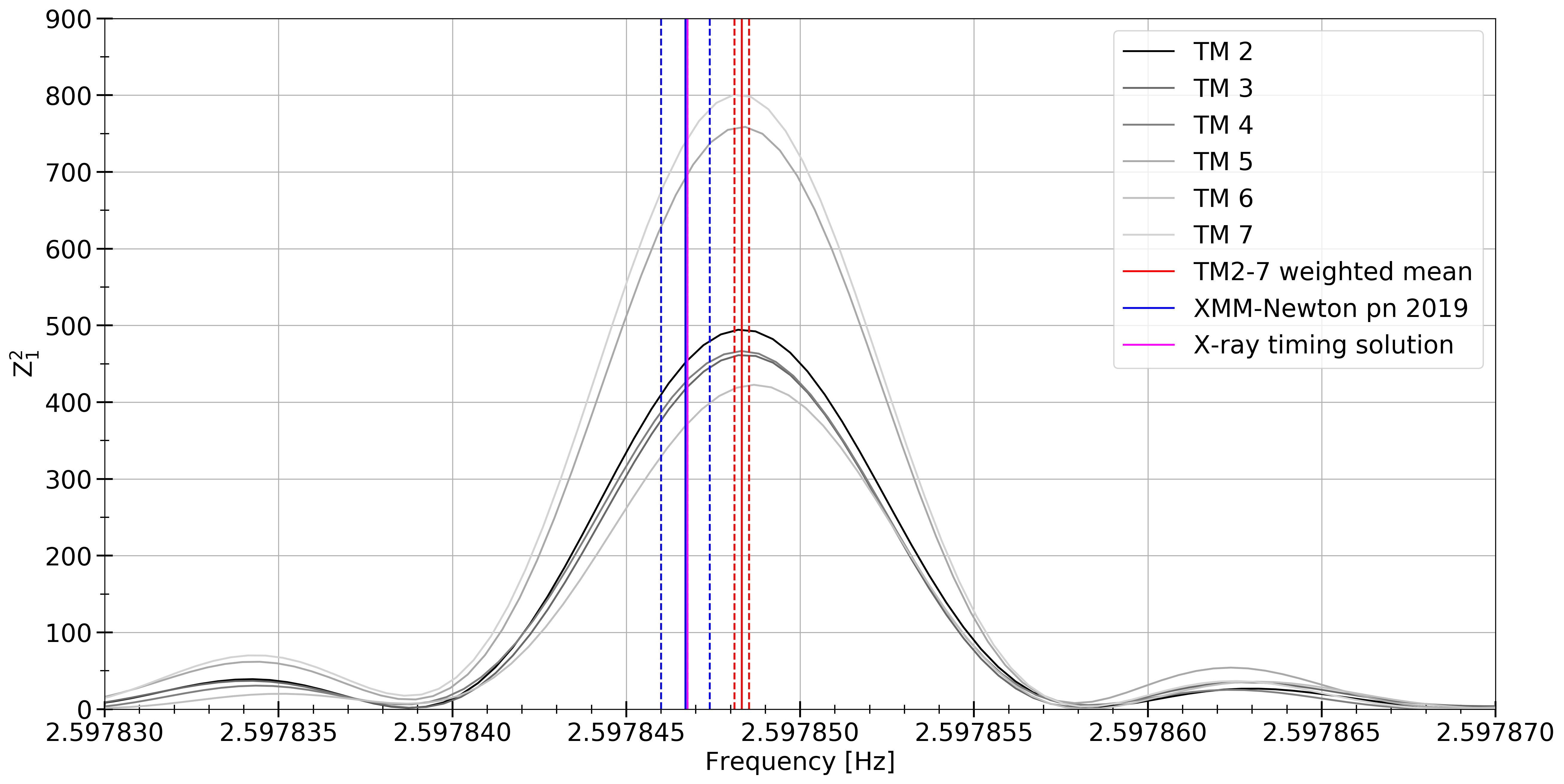}}
\caption{Period search among \ero and \xmmn data. Shown is the $Z^2_1$ statistics per \ero TM (identified with different grey nuances) and the most likely periods from \ero (red vertical line, weighted mean, dashed lines indicate $1\sigma$ confidence interval), the simualtaneous \xmmn data (blue vertical lines) and the extrapolated period using the ephemeris given by \cite{lower+20} (magenta)
 \label{f:z2}
}
\end{figure}

\begin{figure*}[t]
\includegraphics[width=\textwidth]{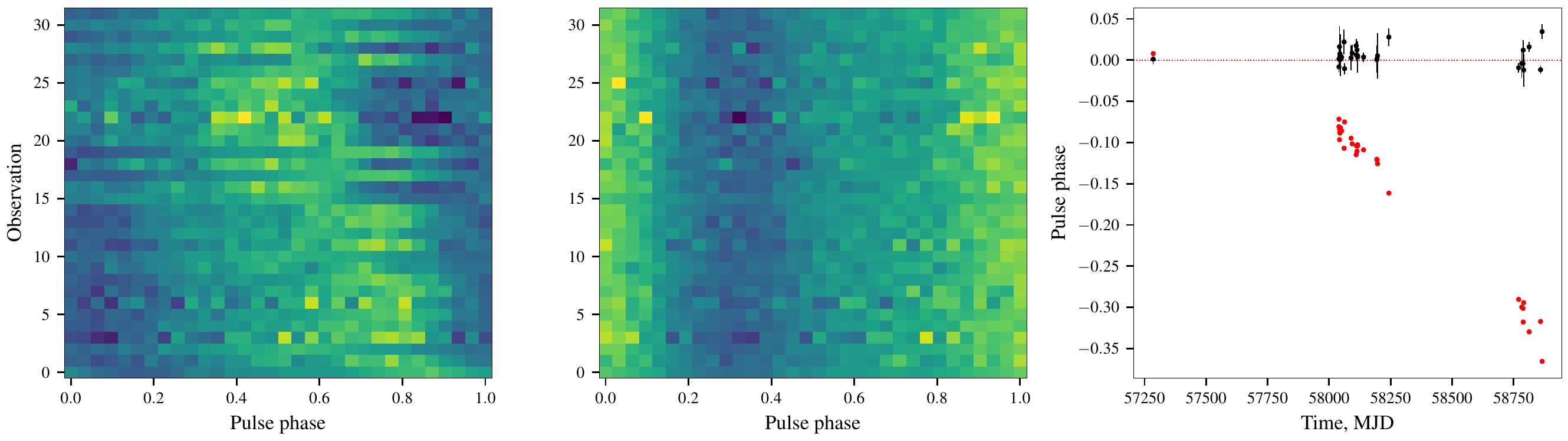}
\caption{Phaseograms showing pulse profiles in 0.3-2\,keV for individual data groups listed in Table~\ref{t:xraytoa} folded using \cite{johnston_kerr18} (left) and the  updated ephemeris (middle) obtained in this work. The phase residuals to the best-fit solution (black symbols with error bars), and for the \cite{johnston_kerr18} solution (red, zero phase corresponds to the phase of the radio peak, errors omitted for clarity) are also shown in the right panel.}
\label{f:xtoas}
\end{figure*}

\subsection{\nus\label{s:nustarreduction}}

To improve hard band coverage for the spectral analysis we included in the analysis the \emph{NuSTAR} observation of the source listed in Table~\ref{t:obslog} and previously reported by \cite{arumugasamy+18} who obtained the data simultaneously with the 2015 \xmmn observation. To reduce the data, we followed the same procedures as described by \cite{zharikov+21}, who used the same observation ion their analysis. In particular, the raw data was reprocessed using the \emph{nuproducts} task from HEASOFTv6.28 and calibration files v20210202. The source spectrum was extracted from a circular region centred on the source with radius of 30\arcsec\ (independently for each telescope module), while the background from a nearby source-free region with radius of 100\arcsec, located approximately at the same location as used by \cite{zharikov+21}.
The extracted spectra for both telescope units were grouped to contain at least 25 counts per energy bin and modeled together with other spectra from other instruments as described in Sect.~\ref{s:avspec}.

\subsection{\nic\label{s:nicerreduction}}

To improve pulsar ephemeris and enable phase-resolved analysis we also analysed \nic observations of the source. 
We retrieved all \nic observations of the source obtained up to the time of the analysis (March 2021). The \nic master catalogue provided by HEASARC lists 93 observations obtained between Oct 13, 2017, and Apr 9, 2020, with an exposure time larger than 0. In addition, the 2015 and 2019 \xmmn observations were included in the analysis. The 2001 observation was not included due to the large time gap between 2001 and 2015 without any sufficiently long X-ray observation. 
Considering that the shapes of X-ray lightcurves are energy dependent \cite[see Sect.~\ref{s:pfold} and papers by ][]{deluca+05,arumugasamy+18}, we only used photons in the range $0.3-2$\,keV where the responses of \nic and \xmmn are similar, to avoid possible apparent distortion of the pulse profiles due to differences in the energy response of the two instruments. 

All \nic observations listed in Table~\ref{t:xraytoa} were reduced using the current set of calibration files and standard screening
criteria\footnote{\url{https://heasarc.gsfc.nasa.gov/docs/nicer/nicer_analysis.html}}. The arrival times in the resulting event lists were then corrected to the solar system barycentre and merged in several groups corresponding to observations performed close in time (i.e.~separated by gaps of at most two days), as summarised in Table~\ref{t:xraytoa}. 

\section{Analysis and results \label{s:results}}

\subsection{Timing analysis \label{s:timing}}

We used the observation of the pulsar to test the relative timing accuracy of the individual \ero TMs. We searched for the pulsar period, using events with energies between 0.3-2.0~keV, in the \xmmn and individual \ero TM exposures applying the $Z^2_n$ test \citep{buccheri+83}. Confidence levels on the frequency of the highest $Z^2_1$ peak were estimated by maximum likelihood \citep[e.g.][]{fisher1990statistical}. The results are listed in Table~\ref{t:per} and illustrated in Fig.~\ref{f:z2}. The $Z^2$ values for TMs 5 and 7 are higher than for the other TMs due to the much higher count rates in these modules. The table also lists the reference periods that are derived by extrapolating recent radio and gamma-ray ephemerides to the date of the joint \ero/\xmmn campaign \citep{ray+11, johnston_kerr18, lower+20}\footnote{We thankfully acknowledge the use of the Lower et al.~updated parameter file, which includes data from the UTMOST program prior to publication.}. Excerpts of the ephemeris parameter files of those three timing solutions are listed in the appendix (Tab.~\ref{t:ephlit}).

\begin{table}[t]
\caption{Results of the period search
\label{t:per}}
\centering
\begin{tabular}{lr}
\hline\hline
Reference & \multicolumn{1}{c}{Period}\\
 & \multicolumn{1}{c}{(ms)} \\
\hline
TM2 & 384.93396(7)\\
TM3 & 384.93396(7)\\
TM4 & 384.93395(7)\\
TM5 & 384.93397(7)\\
TM6 & 384.93399(7)\\
TM7 & 384.93394(7)\\
TM0 & 384.93394(4)\\
\xmmn & 384.93418(10)\\
Ray+11\tablefootmark{a}\,({\it Fermi} {$\gamma$-ray}) & 384.93417645(3)\\
JK18\tablefootmark{b}\,(radio) & 384.93417631(7)\\
Lower+20\tablefootmark{c}\,(radio) & 384.934181690(9) \\
\hline
\end{tabular}
\tablefoot{Numbers in parenthesis indicate uncertainties in the last digits. The periods at the epoch of the observation were extrapolated using the following references:
\tablefoottext{a}{\cite{ray+11}};
\tablefoottext{b}{\cite{johnston_kerr18}};
\tablefoottext{c}{\cite{lower+20}}.} 
\end{table}

The periods found from the six \ero TMs agree well with each other, but individually and jointly (average period) deviate from the simultaneous \xmmn result and from the radio and $\gamma$-ray references. The relative deviation, $(P_{\rm ero}-P_{\rm ref})/P_{\rm ref}$ is $(-6.23, -6.14, -6.14, -6.28) \times 10^{-7}$, respectively from the simultanous \xmmn and extrapolated periods from \cite{ray+11}, \cite{johnston_kerr18}, and \cite{lower+20} ephemerides (referred to as JK18 and Low20 in Table~\ref{t:per}). 

Part of the deviation between \ero and the external references lies in the constantly growing ``clock drift'' between the central SRG quartz and the UTC references, which is of the order of 12\,ms\,day$^{-1}$ (see Appendix~\ref{s:timeshift}, for details). The clock drift accounts for a relative deviation of $-1.4 \times 10^{-7}$, indicating that remaining timing calibration uncertainties are to be found in the \ero time system. 

In the absence of an absolute time reference for \ero, we decided to use a local phase convention further in this analysis. Phase zero was defined as the phase of the maximum count rate in a phase-folded X-ray lightcurve in the energy band $0.3 - 2.0$\,keV. To generate phase-folded lightcurves and phase-binned spectra for both \xmmn and \ero, we used the mission-specific period and sampled the lightcurve into 20 phase bins. For \ero, we used the weighted average period of the six TMs as given in Table~\ref{t:per}, $384.93394(4)$\,ms.
To fold the lightcurves in phase, we define the phase of X-ray maximum by fitting the sum of the first and the second harmonic sine functions,
\begin{displaymath} 
R(\varphi) = a \sin(2\pi \varphi + b) + c  \sin(4 \pi \varphi + d) + e
\end{displaymath}
\noindent where $\varphi$ is the initial arbitrary phase. 
This model represented the observed data well, indicated by the reduced $\chi^2_\nu$ taking on values between 0.78 to 1.41 for 15 degrees of freedom. 

The time that corresponds to the maximum in the best-fit was subtracted from the input data to then generate the final phase-folded lightcurves in different energy bands. The same procedure was adopted for the EPIC-pn data.

\cite{deluca+05} established in their analysis of the first \xmmn data obtained in 2001 a phase relation between the X-ray maximum and the radio pulse. They found that the radio pulse occured $0.25 \pm 0.05$ phase units later than the X-ray maximum, where the latter was defined as the maximum bin in the phase-folded X-ray lightcurve. We were interested whether we would find the same phase relation in the data obtained with \xmmn in 2015 and 2019, which both could be put on an absolute time scale using the JK18 and Low20 ephemerides. However, the slightly longer period determined by \cite{lower+20} leads to a cycle count difference between the 2015 and 2019 observations of 4.8 spin cycles with respect to the JK18 timing solution, prevent us from determining the phase offsets between X-ray radio pulse. 

\begin{figure}[t]
\resizebox{\hsize}{!}{\includegraphics{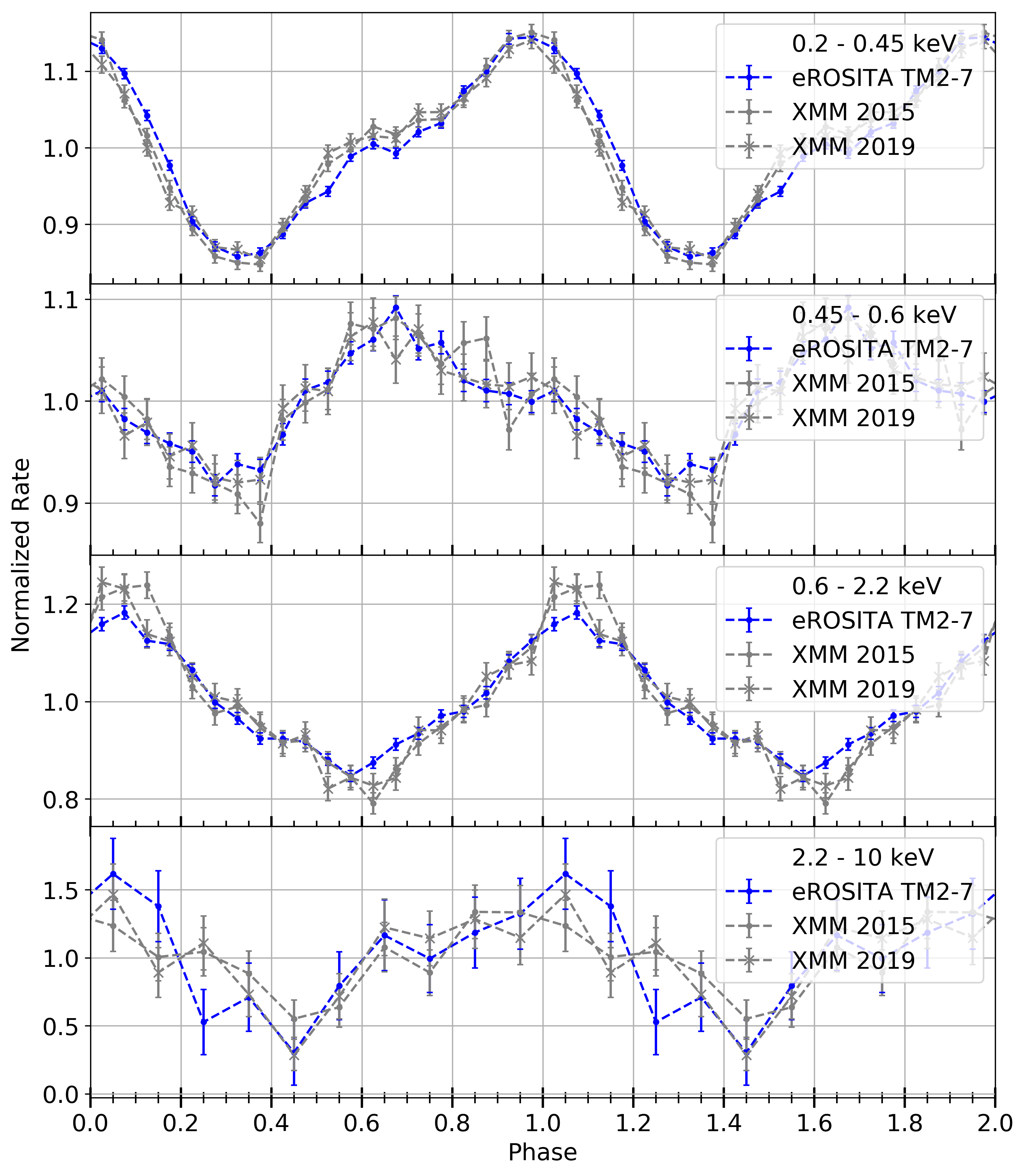}}
\caption{Background subtracted, \ero and \xmmn\ pulse profile in different energy bands. The original data were averaged into 20 phase bins at soft and medium energies and averaged into 10 phase bins at hard energies. The lightcurves are shifted to a common phase, where phase zero is defined as the maximum of the main pulse (see the text; for details).
\label{f:lcs}
}
\end{figure}

\subsubsection{Timing solution of \xmmn and \nic \label{s:timingsolution}}

To resolve the issue we generated our own long-term ephemeris based on X-ray data alone. In particular, \nic and \xmmn data reduced as described in Sects.~\ref{s:xmmreduction} and \ref{s:nicerreduction} were used for this analysis. As already mentioned, all events from both instruments were grouped in several groups separated by gaps not larger than two days as defined in Table~\ref{t:xraytoa}.
Folding events in each group separately using the ephemeris by \cite{johnston_kerr18} reveals that the X-ray pulse phase initially appears constant but then deviations start to grow, eventually reaching $\sim0.35$ phase (see Fig.~\ref{f:xtoas}). That is, there are significant deviations from this timing solution although the pulse cycle counting appears to be retained, so we base our further analysis on this solution. 

It is interesting also to note that the pulse phase of the X-ray peak for the first observation (closest in time to radio data, but still $\sim614$\,d after the end of the formal validity period for radio ephemerides) is consistent with the expected phase of the radio peak. This is inconsistent with the findings by \cite{deluca+05} who found a $\sim$0.25 phase shift between radio and X-rays and suggests either that radio ephemerides break already for the first observation in our set (i.e.~2015 \xmmn observation), or that the offset between radio and X-ray peaks estimated by \cite{deluca+05} has been estimated incorrectly. To determine which is the correct cause one would need to obtain a proper solution including both radio and X-ray data covering the same observation period. Here we focus, however, only on the X-ray data to properly align the \nic and \xmmn data. 

To improve the estimates of local frequency and frequency derivative values we first conducted a $Z^2_3$ search \citep{buccheri+83} around the prediction based on \cite{johnston_kerr18} for the period covered by X-ray data, i.e.~assuming a zero epoch of MJD~57284 (TDB time system). For this search we concatenated event lists from all observations, and maximized the value of the $Z^2_3$ statistic calculated using the \textit{Stingray} \citep{2019ApJ...881...39H} software package, starting from the initial values. The updated solution corresponding to the maximal statistic value largely eliminates the observed phase drift of the X-ray pulses and allows to obtain a high quality template pulse profile by averaging all observations. This, in turn, allows to determine accurate pulse times of arrival (TOAs) for individual data groups and to estimate a final ephemeris using the proper phase-connection. To do that, events in each group were folded separately using the start time of a given interval as a folding epoch and local frequency and frequency derivative estimates obtained above. The average template pulse profile was then directly fitted to the resulting profiles to determine the phase shifts between them (we also allowed for scaling in intensity of the template). The phase shift was then converted to a time shift using the initial estimate of the local pulse frequency and added to folding epoch to determine TOAs for each group listed in Table~\ref{t:xraytoa}.

The observed TOAs were then modeled assuming a constant spin-down rate, i.e.~using the Taylor expansion in phase~$\phi$ of the form:
\begin{displaymath}
\phi(t)=\phi_0+\nu_0(t-t_0)+\frac{\dot{\nu}_0}{2}(t-t_0)^2
\end{displaymath}
\noindent where $t_0=$~MJD~57284 is our reference epoch, and $\nu_0,\dot{\nu}_0$ are the spin frequency and frequency derivative at that time. Considering the potential presence of timing noise that effectively increases the uncertainty of individual TOA estimates, we used for the final fit the nested sampling Monte Carlo algorithm \texttt{MLFriends} \citep{2014arXiv1407.5459B,2017arXiv170704476B} implemented in the \texttt{UltraNest}\footnote{\url{https://johannesbuchner.github.io/UltraNest/}} package to derive posterior probability distributions and the Bayesian evidence for individual parameters. The best-fit residuals with regard to our final solution are presented in Fig.~\ref{f:xtoas}, and correspond to $\nu_0$=2.5978943932(1)\,Hz, $\dot{\nu}_0=3.70791(2)\times10^{-13}$\,Hz\,s$^{-1}$, where parameter values and uncertainties are derived from their final posterior probability distributions. We emphasise that the solution above is an approximation that is valid for the period from MJD~57284 to MJD~58866, and a joint fit of X-ray and radio arrival times over a longer baseline is required to determine a final timing solution for the pulsar.

\begin{table}[t]
\caption{Pulsed fractions ($p_{\rm f}$) and phases of pulse maxima ($\varphi_{\rm max}$) in different energy bands for the 2019 \ero/\xmmn campaign
\label{t:lcp}}
\centering
\begin{tabular}{c|lr|lr}
\hline\hline
Energy band & \multicolumn{2}{c|}{\ero} & \multicolumn{2}{c}{\xmmn 2019} \\
\cline{2-3}\cline{4-5}
(keV) & $p_{\rm f}$ & $\varphi_{\rm max}$ & $p_{\rm f}$ & $\varphi_{\rm max}$ \\ 
\hline
1: $0.20-0.45$ & 0.143(5) &0.98 & 0.143(8) & 0.96 \\
2: $0.45-0.6$ & 0.087(8) &0.66 & 0.079(17) & 0.64 \\
3: $0.60-2.2$ & 0.165(9) &0.06 & 0.205(19) & 0.06 \\
4: $2.2-10$ & 0.68(22) &0.05\tablefootmark{a} & 0.67(12) & 0.95\tablefootmark{a} \\
\hline
\end{tabular}
\tablefoot{
\tablefoottext{a}{Due to low statistics, the indicated values were not fitted but estimated.}}
\end{table}

\begin{figure}[t]
\resizebox{\hsize}{!}{\includegraphics{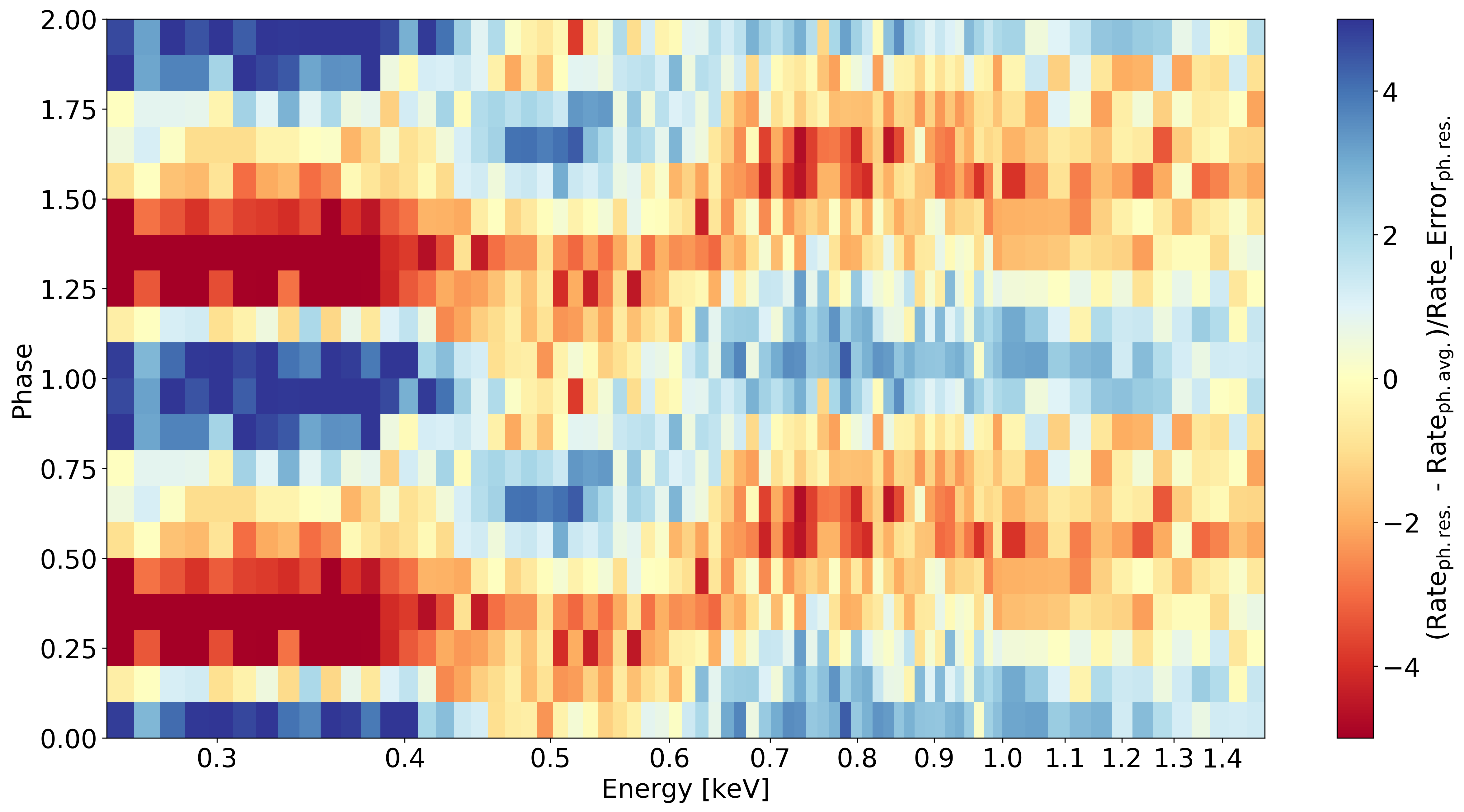}}
\caption{Trailed spectrogramme of the \ero pulse profile, with phase and energy along the $y$ and $x-$axes, respectively. The mean spectrum was subtracted from the original data. Color thus indicates a positive or negative deviation from the mean. Excess values were normalised to their errors. The pulse profile is shown twice for better visibility.
\label{f:trail}}
\end{figure}

\subsubsection{Phase-folded lightcurves \label{s:pfold}}

Using the timing solution from Sect.~\ref{s:timingsolution}, we created phase-folded pulse profiles in various energy bands and binned into 20 phase bins (Fig.~\ref{f:lcs}). Phase zero is defined as the time of the X-ray maximum in the energy band $0.3-2.0$\,keV, which was determined by the harmonic fit described above. As noticed previously \citep[][]{deluca+05,arumugasamy+18}, the phase of X-ray maximum is energy-dependent. The energy bands of the lightcurves in Fig.~\ref{f:lcs} were chosen to highlight the variability of the main spectral components: the cool blackbody, the region where the absorption feature occurred, the hot blackbody, and the power-law tail. Our analysis reveals that the shapes of the lightcurves in the various bands and their phase relation were found to be the same in 2001, 2015, and in 2019. In Table~\ref{t:lcp} we list the phase of the X-ray maximum and the pulsed fraction per energy band per mission for the 2019 \ero/\xmmn campaign. The pulsed fractions were calculated from the maximum and minimum count rates in the phase-folded lightcurves using
\begin{displaymath}
PF = \frac{R_{max}-R_{min}}{R_{max}+R_{min}} 
\end{displaymath}
while the error was propagated from the errors of the maximum and minimum count rates.

The low energy band (1) shows a slow increase and a fast decrease towards a minimum at phase 0.35. At all occasions it shows a shoulder at phase 0.6, at the same phase when the minimum of the intermediate band (3) occurrs, which traces the hotter blackbody. At this energy (band 3) the lightcurve is almost symmetric, as one would expect from simple foreshortening and projection of a well-behaved, i.e.~symmetric heated spot on the NS surface. The phase offset between the soft band (1) and the intermediate band (3) is significant. The hot blackbody peaks at phase 0.06, i.e.~0.1  phase units later than the cold blackbody (see also Sect.~\ref{s:phrspec} below for the results of phase-resolved spectroscopy). The intermediate band (2) behaves  much different than the blackbody-dominated bands (1) and (3). It peaks at about phase 0.65, i.e.~at the shoulder of band (1), and has a much shallower variability amplitude. Although this band was chosen to somehow trace the absorption feature, the observed lightcurve shape, which differs strongly from bands (1) and (3), is not due to the absorption feature. This weak feature has a much smaller impact on the shape of the lightcurves compared to the continuum emission processes. The presence of the shoulder in band (1) which coincides with the maximum in band (2) hints to a more complex emission region than that of a symmetric cold spot (cold compared to the hot blackbody component). Either emision is from one structured cold region or from a third region with a temperature not  much different from that causing the main maximum in band (1). 

The hard band is photon starving and the derived parameters have large uncertainties. Its low energy boundary was chosen such that the contribution of the hot blackbody was below 3\%. At the S/N and the phase resolution achieved with our observations it may be described with just a more or less symmetrical bright phase and a main minimum which occurrs are phase 0.40-0.45. The maximum occurrs at around phase 0.9-1.0.

It is worth noting that phase-smearing of \ero data cannot be recognised. The frame time of the \ero cameras is 50 msec, which corresponds to $<8$ independent phase bins per spin revolution. The EPIC-pn small window mode has a frame time of 5.7 msec (67 independent phase bins). As shown in Fig.~\ref{f:lcs} and Table~\ref{t:lcp} the pulsed fractions of the lightcurves in band (1) are the same for \ero and \xmmn while there is only little degradation of the pulsed fraction seen in \ero in bands (2) and (3) despite the lower sampling.

In previous studies involving data from \xmmn energy-resolved light curves were generated as well \citep[][their figures 6 and 9, respectively]{deluca+05,arumugasamy+18}. Both used slightly different energy passbands and 10 phase bins for all light curves. All curves in the soft band, dominated by the cool blackbody, are very similar with a shoulder on the rising branch. Our curve for band 3, dominated by the hot blackbody, is a bit dissimilar compared to the others. Ours appears rather symmetric with a small shoulder on the descending branch which appears more pronounced in the other works. The pulsed fraction is highest in all studies in band 4, dominated by the power law. The curve shown by \cite{arumugasamy+18} looks nevertheless different from ours. It has one bin at their phase 0.45 with a rate as high as during the main hard pulse, i.e. at phase $\sim$0.1, which we did not find, neither in the two data sets from \xmmn, nor in the \ero data. 

A complementary view on the energy- and phase-dependent variability is given in Fig.~\ref{f:trail} which shows the photons in the soft energy range ($0.3 - 1.5$\,keV) arranged as a trailed spectrogam with the mean subtracted. 
It shows that the light-curve maxima below 0.45 keV and above 0.7 keV are stable in phase. It also shows a transition region between 0.45 and 0.7 keV with a strongly variable phase of maximum emission. Above an energy of $\sim$0.7\,keV the maximum phase is apparently stable. Our figure can be compared with the corresponding Figure 10 from \cite{arumugasamy+18} showing the same features albeit with a smaller energy range covered.

\begin{figure}[t]
\resizebox{\hsize}{!}{\includegraphics{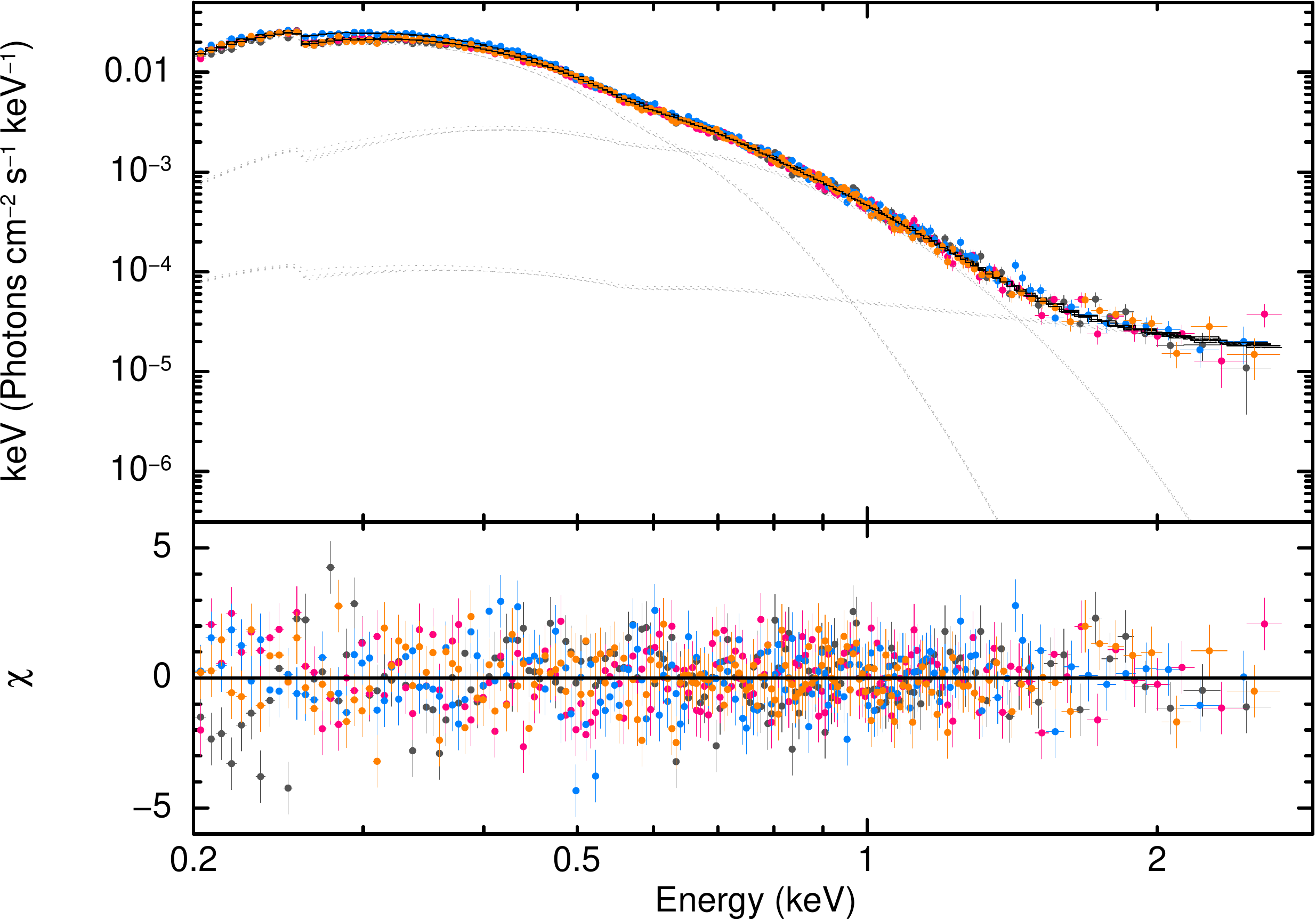}}\\[0.5cm]
\resizebox{\hsize}{!}{\includegraphics{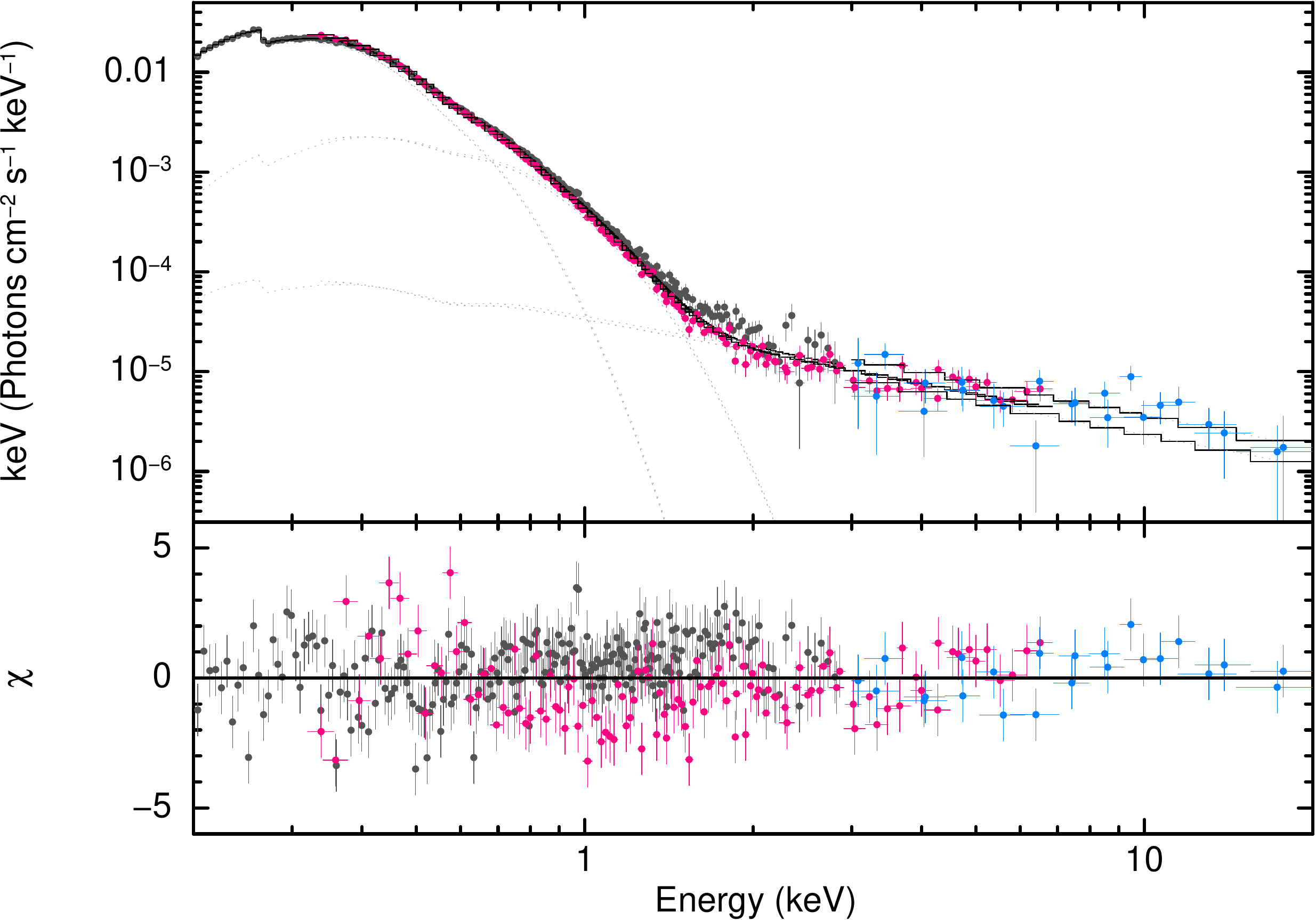}}
\caption{Results of the phase-averaged spectral fitting. {\it Top.} We show the \ero spectra of detectors 2, 3, 4, and 6 fitted simultaneously by the model of Table~\ref{t:avspec} and a fixed power-law index of $\Gamma=1.98$. {\it Bottom.} Simultaneous fit of \ero, \xmmn, and \nus (dark grey, magenta, and blue data points, respectively). The best-fit parameters are as in Table~\ref{t:avspec}.
\label{f:avspec}
}
\end{figure}

\begin{table*}[t]
\small
\caption{Results of the phase-averaged spectral modeling
\label{t:avspec}}
\centering
\begin{tabular}{lrrrrrrrrrrrr}
\hline\hline
ID\tablefootmark{$\dagger$}& $\chi^2_\nu$\,(dof) & $N_{\rm H}$\tablefootmark{(a)} & \multicolumn{2}{c}{Edge} & \multicolumn{1}{c}{$ kT_1$} & \multicolumn{1}{c}{$R_1$\tablefootmark{(b)}} & \multicolumn{1}{c}{$kT_2$} & \multicolumn{1}{c}{$R_2$\tablefootmark{(b)}} & \multicolumn{1}{c}{$\Gamma$} & \multicolumn{1}{c}{$\epsilon$} & \multicolumn{1}{c}{$\sigma$} & \multicolumn{1}{c}{$f_{\rm X}$\tablefootmark{(c)}} \\
\cline{4-5}
 & & & ($E$; eV) & \multicolumn{1}{c}{$\tau$} & \multicolumn{1}{c}{(eV)} & \multicolumn{1}{c}{(km)} & \multicolumn{1}{c}{(eV)} & \multicolumn{1}{c}{(km)} & & \multicolumn{1}{c}{(eV)} & \multicolumn{1}{c}{(eV)} & \\
\hline
$[1]$ & 1.6\,(245) & $1.00_{-0.05}^{+0.05}$ & $-$ & $-$ & $76.6_{-1.0}^{+1.0}$ & $6.44_{-0.18}^{+0.21}$ & $142_{-4}^{+4}$ & $0.56(6)$ & $1.94_{-0.25}^{+0.27}$ & $592_{-3}^{+3}$ & $79_{-5}^{+5}$ & $1.1140(26)$ \\ 
$[2]$ & 3\,(391) & $1.22_{-0.03}^{+0.03}$ & $-$ & $-$ & $73.4_{-0.2}^{+0.3}$ & $7.3_{-1.1}^{+1.2}$ & $135.5_{-0.5}^{+0.3}$ & $<0.7$ & $1.87_{-0.07}^{+0.08}$ & $588.5_{-1.9}^{+2.0}$ & $75_{-3}^{+3}$ & $1.1064(15)$\\ 
$[i]$ & 1.1\,(243) & $1.68_{-0.15}^{+0.16}$ & $260.0_{-2.0}^{+2.0}$ & $0.32_{-0.03}^{+0.04}$ & $64.8_{-2.0}^{+2.1}$ & $11_{-4}^{+5}$ & $120_{-3}^{+3}$ & $1.1_{-0.5}^{+0.5}$ & $2.50_{-0.20}^{+0.3}$ & $571_{-4}^{+4}$ & $67_{-7}^{+8}$ & $1.113(3)$\\ 
$[ii]$ & 1.6\,(389) & $1.66_{-0.09}^{+0.10}$ & $265.6_{-2.6}^{+2.0}$ & $0.360_{-0.016}^{+0.017}$ & $67.2_{-2.0}^{+1.7}$ & $10_{-3}^{+4}$ & $126_{-3}^{+3}$ & $0.9_{-0.4}^{+0.5}$ & $1.98_{-0.09}^{+0.10}$ & $569_{-4}^{+3}$ & $80_{-11}^{+7}$ & $1.1122(26)$\\ 
$[iii]$ & 1.1\,(244) & $1.55_{-0.12}^{+0.13}$ & $261.0_{-3}^{+2.0}$ & $0.29_{-0.03}^{+0.03}$ & $ 66.8_{-1.8}^{+1.8}$ & $10_{-4}^{+4}$ & $125_{-2}^{+3}$ & $1.0_{-0.4}^{+0.4}$ & $1.98$\tablefootmark{$\star$} & $572_{-4}^{+4}$ & $71_{-7}^{+8}$ & $1.1176(26)$\\ 
\hline
\end{tabular}
\tablefoot{The composite phenomenological model fitted to the data consists of an absorbed double blackbody plus power-law continuum, modified by a Gaussian absorption line, with $[i, ii, iii]$ and without $[1, 2]$ an edge component. The latter is introduced to take into account strong residuals at low energies, predominantly in \ero data (see the text for details). Errors are $1\sigma$ confidence levels. 
\tablefoottext{$\dagger$}{We either perform a simple fit of TM8 [1, $i$, $iii$], or a simultaneous fit of the five data sets: TM8, \xmmn pn (2015, 2019), \nus (both telescope units) [$2$, $ii$]. In [$iii$] we re-fit TM8 with a fixed power-law slope $\Gamma=1.98$, as found in the joint fit. The renormalisation factors of the simultaneous fit [2], with respect to the TM8 dataset, are $1.0267(27)$, $1.0058(29)$, $1.04_{-0.13}^{+0.14}$, $0.68_{-0.12}^{+0.13}$, respectively, for pn (2015), pn (2019), and both \nus modules. The same factors are $0.951(4)$, $0.933(4)$, $1.03_{-0.13}^{+0.15}$ and $0.66_{-0.11}^{+0.13}$ for fit [$ii$].}
\tablefoottext{a}{The column density is in units of $10^{20}$\,cm$^{-2}$.}
\tablefoottext{b}{The radiation radius at infinity for each blackbody component is computed assuming a distance of 288\,pc.}
\tablefoottext{c}{The observed model flux is in units of $10^{-11}$\,erg\,s$^{-1}$\,cm$^{-2}$ in energy band 0.2--12 keV.}
\tablefoottext{$\star$}{Parameter kept frozen during fitting.}}
\end{table*}

\subsection{Spectral analysis \label{s:spec}} 

The analysis of the \ero data is based on source and background spectra extracted from regions as described in Sect.~\ref{s:obs}, together with the respective response matrices and ancillary files created for each detector with the eSASS task {\it srctool}. We restricted the analysis to GTI-filtered photons with energy within 0.2\,keV and 5\,keV, beyond which the source signal-to-noise ratio becomes insignificant. The energy channels of each TM spectrum were regrouped with the HEASOFT task {\it grppha} to avoid a low ($<50$) number of counts per spectral bin.

For the EPIC-pn data (Sect.~\ref{s:xmmreduction}), we used the SAS tasks {\it rmfgen} and {\it arfgen} to generate the respective RMF and ARF files for the 2015 and 2019 observations. The spectra were rebinned with the SAS task {\it specgroup} to ensure a minimum $S/N$ ratio of 4, while keeping oversampling of the instrument energy resolution within a factor of 3. The EPIC data were analysed within $0.3-7$\,keV, in accordance with the guidelines and calibration status of EPIC-pn in SW mode.

The RGS GTI-filtered event lists were used to extract the source and background spectra in wavelength space using the SAS tasks {\it rgsregions} and {\it rgsspectrum}, while response matrix files were produced with {\it rmfgen}. Only the first-order spectra were analysed. To increase the S/N we co-added the RGS1 and RGS2 spectra of the 2015 and 2019 observations into two stacked data sets using the SAS task {\it rgscombine}; each combined spectrum were then rebinned into 0.165\,\AA\ wavelength channels. The defective channels of the RGS cameras (Sect.~\ref{s:xmmreduction}) were excluded from the spectral fitting. The co-added background and response files in each detector were taken into account in the spectral fitting as usual. The total data set respectively amounts to $1.935(14)\times10^4$ and $2.496(16)\times10^4$ counts ($12-38$\,\AA) in each RGS1/2 camera, of which around 40\% can be ascribed to the background.

To fit the spectra we used \xspec 12.10.1f \citep{arnaud+96}. Unless otherwise noted, the fit parameters were allowed to vary freely within reasonable ranges. Whenever spectra from different instruments were fitted simultaneously, we adopted a renormalisation factor between them to take into account calibration uncertainties (see caption of Tables~\ref{t:avspec} and \ref{t:phrespar}). The photoelectric absorption model and elemental abundances of \citet{wilms+00} were adopted to account for the interstellar material in the line-of-sight. Due to the low absorption towards the target, the choice of abundance table does not significantly impact the results of the spectral fitting.

\subsubsection{The phase-averaged X-ray spectrum of \psr \label{s:avspec}}

The main emission components in the phase-averaged spectrum of \psr are well described in the literature \citep[see, e.g., ][and references therein]{zharikov+21}. We re-do the exercise here because different time, pattern and region selection might lead to slightly different source parameters from missions, whose data were available to previous researchers. The inclusion of the data from \ero may shed new light on the emission model of \psr and has two aspects. Firstly, we are interested in the level of agreement of spectral parameters between \ero and \xmmn to assess the calibration uncertainties, and secondly, given the improved spectral resolution of \ero with respect to  \xmmn some of the spectral parameters might need to be revised. Also, with a fit to the phase-averaged spectrum we intend to pre-determine some of the spectral parameters for the fits to the phase-resolved spectra which of course have a lower signal-to-noise ratio per spectral bin. 

The basic model we applied to the data consists of the sum of a hot and a cold blackbody, superposed by an absorption feature that is modeled as a Gaussian, plus a power-law hard tail, everything modified by interstellar absorption. In \xspec terminology this model is written as {\tt tbabs((bbodyrad+bbodyrad+powerlaw)gabs)}. 
Until now this was sufficient to describe the \nus and \xmmn spectra, but the model leaves strong residuals below 0.3\,keV in \ero data, which indicate the presence of an additional feature of as yet uncertain nature. If the residuals are not taken into account in the spectral fitting, the corresponding reduced chi-square values are 1.6 and 3, respectively, for the simple fit of TM8 and simultaneous multi-mission fit (245 and 391 degrees of freedom; cf.~fit IDs $[1]$ and $[2]$ in Table~\ref{t:avspec}). 

The possible presence of a second absorption line has previously been reported by \citet{zharikov+21}, when the authors included photons below 0.3\,keV in their analysis of \xmmn data. Here, we tentatively model the low-energy residuals as a multiplicative absorption edge\footnote{In \xspec, {\tt tbabs((bbodyrad+bbodyrad+powerlaw)gabs)edge}, cf.~fit IDs $[i, ii, iii]$ in Table~\ref{t:avspec}.}, which is favoured over another Gaussian absorption for the main following reasons: first, the residuals are close to the low-energy cutoff of the spectra, therefore the line energy and width of the additional Gaussian component are poorly defined; second, due to co-variance of the multiple model components, the fit with two Gaussians leaves both the column density and the parameters of the cold blackbody largely unconstrained.

We first verified the cross-calibration of the six \ero detectors for various combinations of photon patterns. We investigated the stability of the best-fit solutions against whether simple fits of concatenated event lists, or simultaneous fits of individual TM data sets, were used. The results are documented in detail in Sect.~\ref{s:eroccalib}. Based on this analysis, and given the as yet uncertain energy calibration of TM9 as a consequence of the light leaks, we hereafter adopt the TM8 data set and all valid patterns in the joint analysis of \ero with \nus and \xmmn. 

The fit results of the above described phenomenological model are summarised in Table~\ref{t:avspec}. We list for each entry the reduced chi-square $\chi^2_\nu$ and degrees of freedom (d.o.f.), the equivalent hydrogen column density $\nh$ in units of $10^{20}$\,cm$^{-2}$, the edge energy $E$ in eV and absorption depth $\tau$, the temperature of the cold $kT_1$ and hot $kT_2$ blackbody components in eV, the radiation radii $R_1$ and $R_2$ of each component in km (assuming a distance to the source of $d=288$\,pc; \citealt{brisken+03}), the power-law photon index $\Gamma$, the central energy $\epsilon$ of the absorption line and its Gaussian sigma $\sigma$ in eV, and the observed model flux in the $0.2-12$\,keV energy band. In $[i]$ we show the best-fit results of the TM8 data alone; in $[ii]$ are the results of the joint, multi-mission, fit; $[iii]$ shows again the fit results of the \ero TM8 data with the spectral index fixed to the value found for the joint fit, $\Gamma=1.98$. For the joint fit $[ii]$, we keep the absorption depth of the edge component fixed to zero (multiplicative factor of 1) in the \xmmn and \nus spectra, so that the best-fit edge energy and absorption edge are solely determined over the \ero data set.

As long as the residuals at $260-265$\,eV are taken into account, our results are in general agreement with those reported in the literature, in particular \citet[][column G2BBPL in their Table~2]{arumugasamy+18} and \citet[][Table~7]{zharikov+21}. In particular for the joint multi-mission fit $[ii]$, we found consistent parameters for both thermal components, with comparable relative errors of 2\% to 3\% in temperature and within 40\% and 50\% in blackbody normalisation. The inclusion of photons below 0.3\,keV allows for a better determination of the interstellar absorption, here constrained down to 5\% in comparison to 25\% found in \xmmn data alone. Likewise, the characterisation of the absorption feature first reported by \citet{arumugasamy+18} is significantly improved: the relative errors are below 1\% and around 14\%, respectively, in its central energy and Gaussian width. For comparison, \citet{arumugasamy+18} reports errors of 5\% and 30\% on the same parameters. This is a direct result of the improved energy resolution of \ero with respect to \xmmn. Remarkably, our $N_{\rm H}$ value is roughly half that found by \citet{arumugasamy+18} and the Gaussian feature is determined at a higher energy, $\sim570$\,eV instead of $540$\,eV. These two best-fit results are inconsistent (not within the reported $10^{\rm th}-90^{\rm th}$ confidence percentile) with those of \citet{arumugasamy+18}, who also reports a somewhat less steep power-law slope than what is presented here, but in general agree with fit IDs N3 and N4 of \citet{zharikov+21}.

Additionally, we investigated the RGS spectra for the presence of the absorption line. We fit the two stacked RGS1 and RGS2 spectra simultaneously within $12-38$\,\AA. We verified that an absorbed double blackbody model, {\tt tbabs(bbodyrad+bbodyrad)} in \xspec, fits the RGS data well; a power law is not necessary to describe the continuum given the much more narrow energy range of the RGS instruments. For the same reason, we did not fit the column density, instead fixing it to the best-fit value found for the multi-mission spectral fit of Table~\ref{t:avspec}. Although residuals are present around the wavelength of the absorption line ($21-22.5$\,\AA) in the combined RGS1 spectrum\footnote{Due to the defective channels of RGS2 around the wavelength range of interest, the RGS2 spectrum cannot constrain the presence of the narrow line.}, the inclusion of a Gaussian feature is not statistically required. Nonetheless, adding a Gaussian component to the model gives evidence for a narrow feature at best-fit energy of $\epsilon=557_{-15}^{+16}$\,eV and $\sigma=29_{-19}^{+26}$\,eV, with no significant changes to the model continuum. The energy of the feature agrees within the errors with those found in pn data alone and \ero if a 10\,eV systematic uncertainty of EPIC-pn in small window mode is accounted for. 

We note that there is an obvious disagreement between the instruments which reflects the current state of cross-calibration of \ero and \xmmn. While the fit using just \ero data revels a flat distribution of the residuals (upper panel in Fig.~\ref{f:avspec}) we find systematic deviations in such distributions for \ero and \xmmn data (lower panel). In principle, these can be accounted for by the inclusion of systematic errors at a few percent level ($\sim$3-4\%) for both instruments. This would slightly alter the best-fit parameters and increase the reported uncertainties by a factor of $\sim$3. However, this is clearly not an optimal solution to the problem of calibration of the energy scale and effective area for both instruments, which constitutes a separate ongoing effort by the \ede collaboration and will be reported elsewhere. In Section~\ref{s:atmoco} below we focus, therefore, exclusively on the analysis of \ero data reduced as described above to reflect the current state of the instrument’s calibration. On the other hand, the results of the multi-mission fit are included in this paper to a) document the current state of cross-calibration between \ero/\xmmn (EPIC-pn) and b) to improve the constraints on the power-law index which is poorly constrained using \ero data alone.

\begin{figure}[t]
\resizebox{\hsize}{!}{\includegraphics{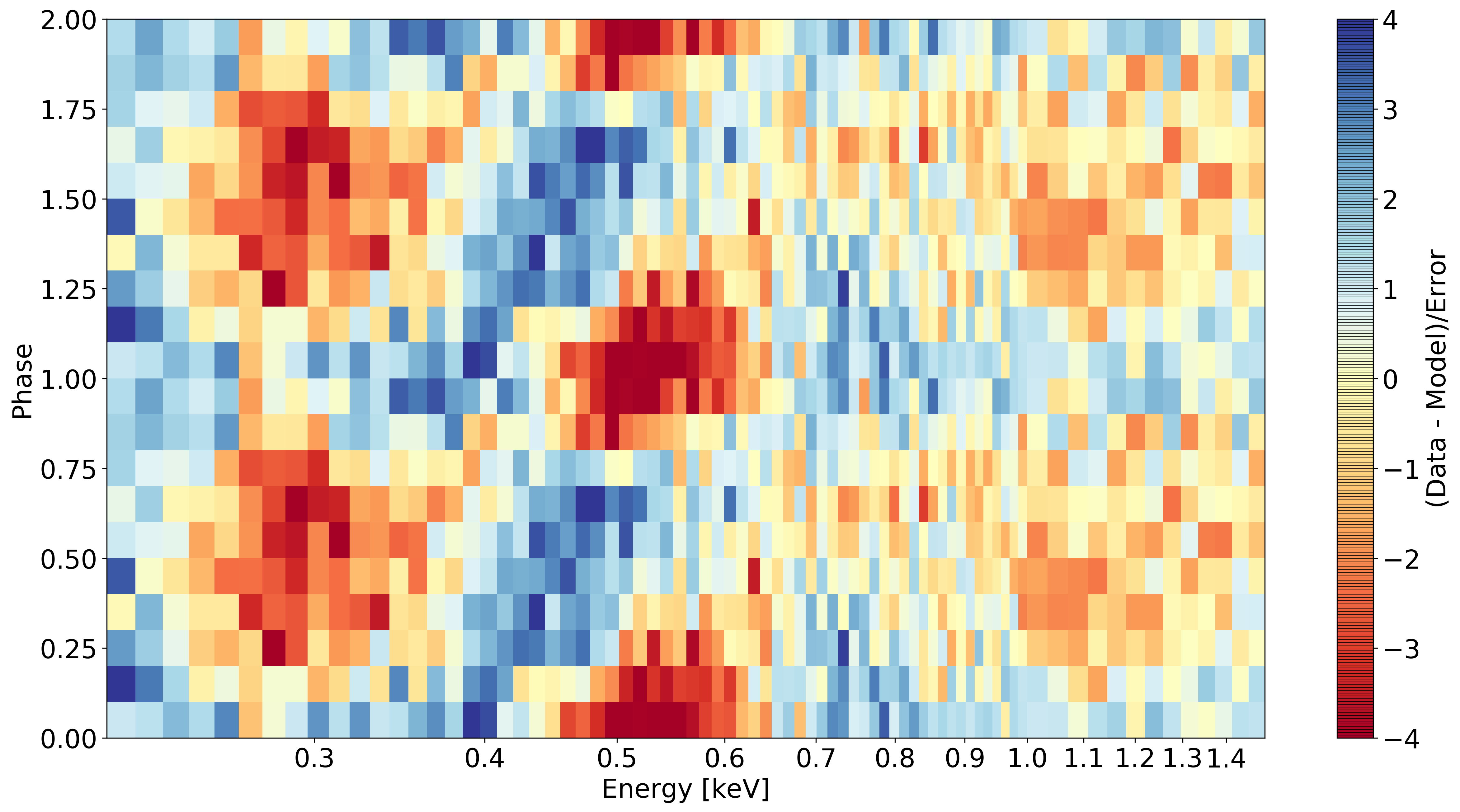}}
\resizebox{\hsize}{!}{\includegraphics{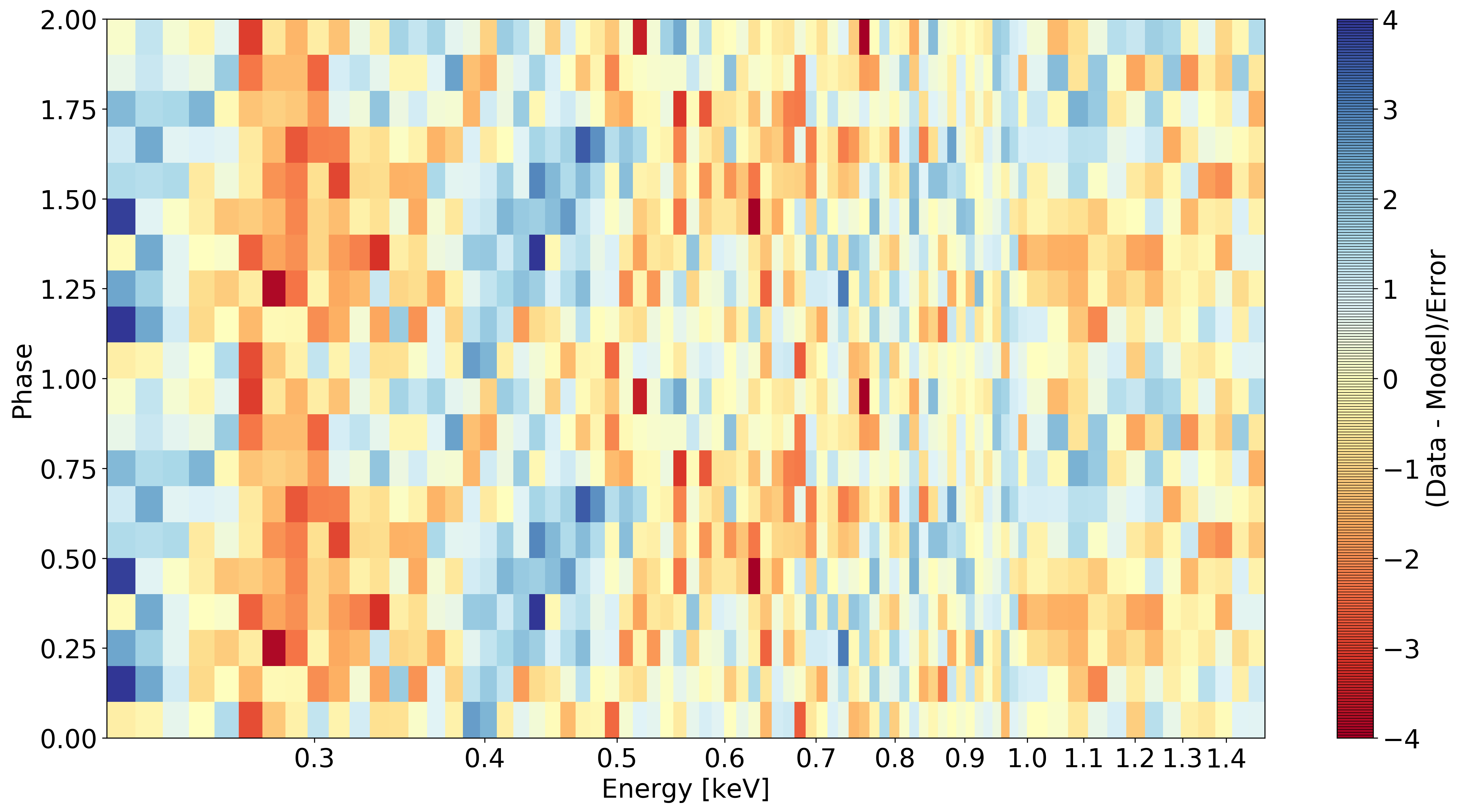}}
\resizebox{\hsize}{!}{\includegraphics{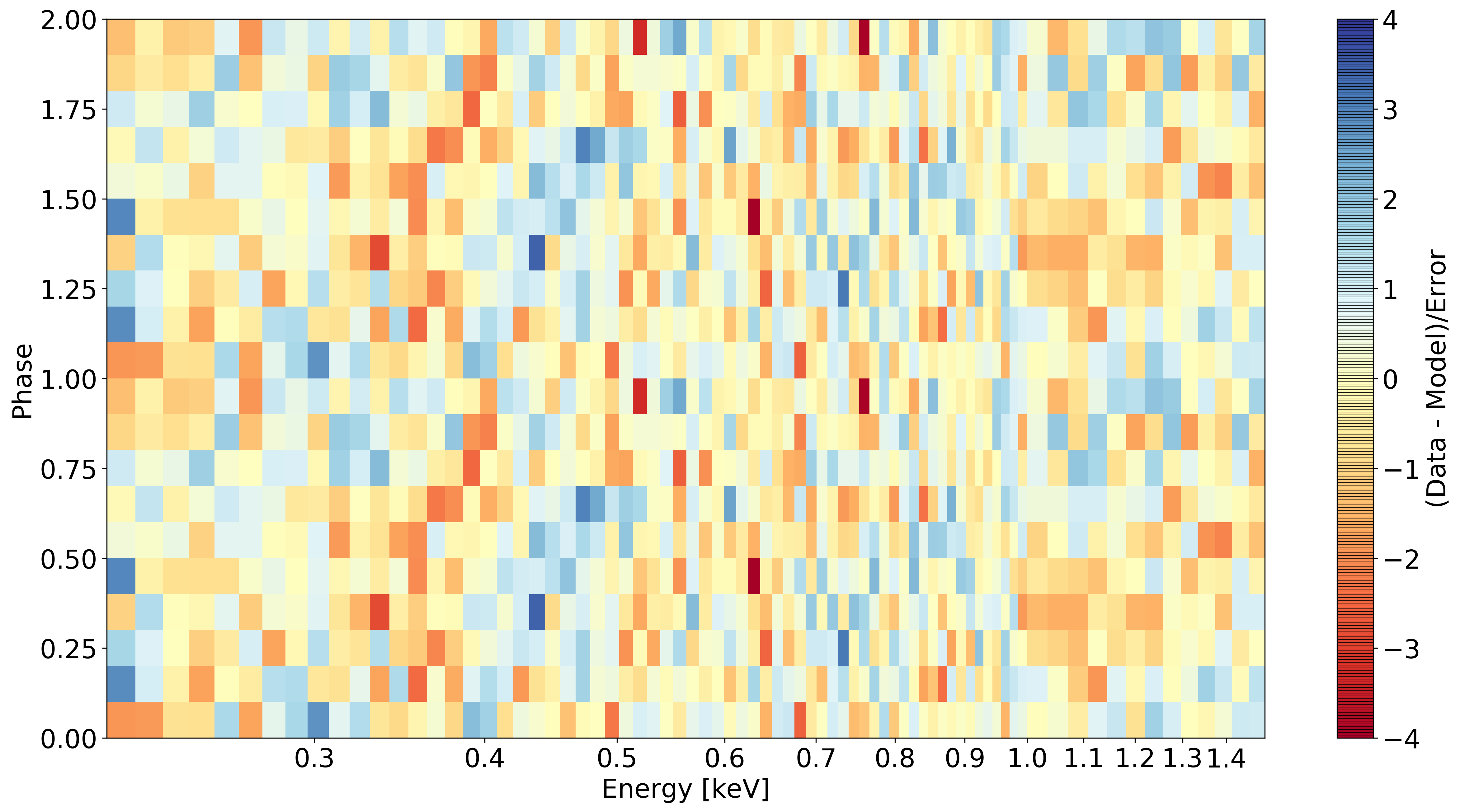}}
\caption{The fit residuals from fitting the TM8 phase-resolved spectra using a 2BBPL (top), a G2BBPL (middle), or a G2BBPLe (bottom) model are shown.
\label{f:phrresid}
}
\end{figure}

\begin{figure}[t]
\resizebox{\hsize}{!}{\includegraphics{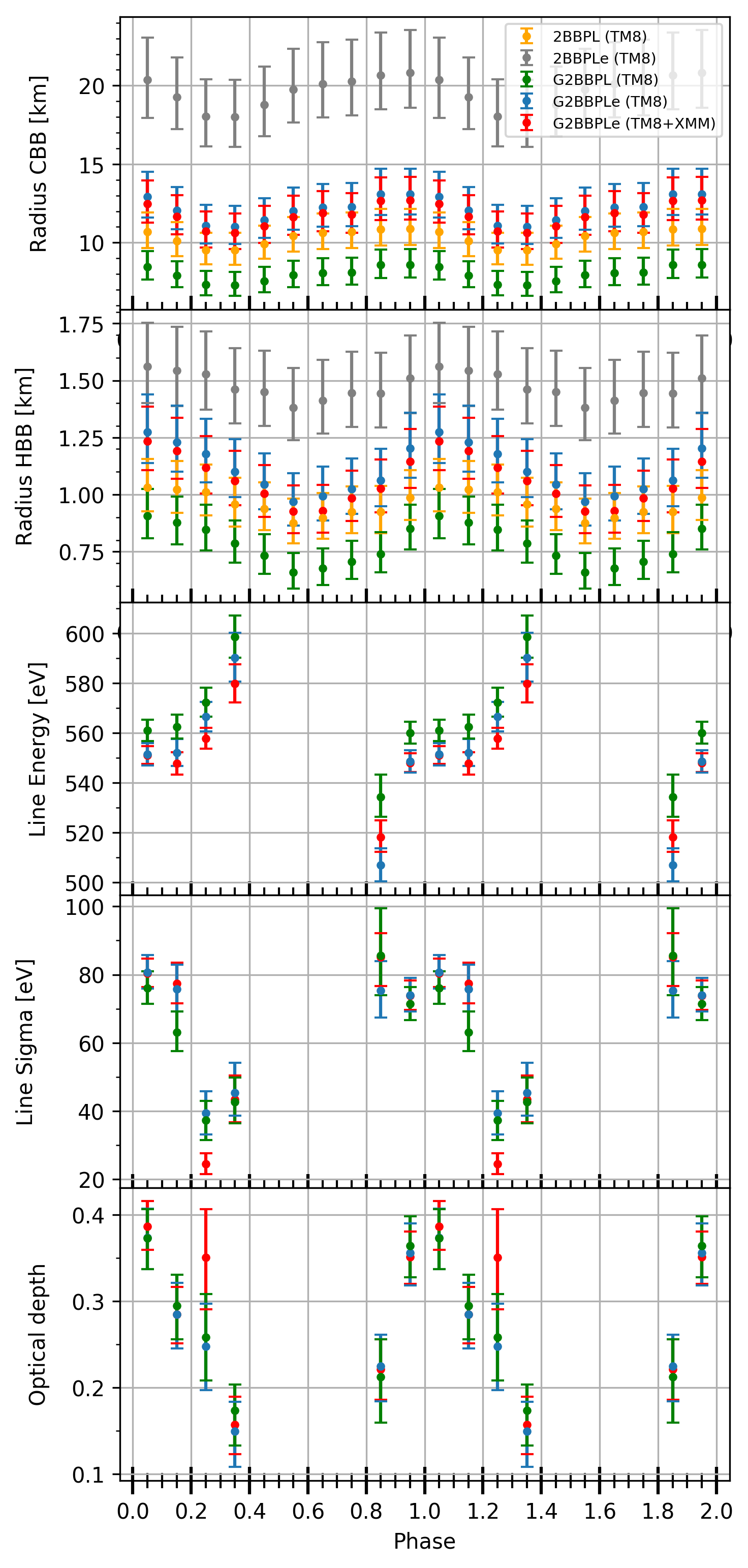}}
\caption{ The results for the phase-dependent parameters, estimated from the phase-resolved fit using a 2BBPL, 2BBPLe, G2BBPL or G2BBPLe model, are depicted. The shown errors are $1\sigma$ confidence levels. 
\label{f:phrparam}
}
\end{figure}

\begin{table*}[t]
\small
\caption{Results of the phase-resolved spectral modeling
\label{t:phrespar}}
\centering
\begin{tabular}{clrrrrrrrrrrr}
\hline\hline
& $\chi^2_\nu$\,(dof) & $N_{\rm H}$\tablefootmark{(a)} & \multicolumn{2}{c}{Edge} & \multicolumn{1}{c}{$ kT_1$} & \multicolumn{1}{c}{$kT_2$} & \multicolumn{1}{c}{$\Gamma$}\\
\cline{4-5}
 & & & ($\epsilon$; eV) & \multicolumn{1}{c}{$\tau$} & \multicolumn{1}{c}{(eV)} & \multicolumn{1}{c}{(km)} & \multicolumn{1}{c}{ } & \\
\hline
TM8 2BBPL & 2.21(1464)  & $1.75^{+0.05}_{-0.05}$ &  &  & $64.5^{+0.5}_{-0.5}$ & $126.8^{+1.3}_{-1.4}$ &  $1.63^{+0.19}_{-0.18}$ \\
TM8 2BBPLe & 1.96 (1462) & $2.66^{+0.12}_{-0.11}$ & $260.3^{+1.5}_{-1.5}$ & $0.512^{+0.04}_{-0.029}$ & $55.1^{+0.8}_{-1.1}$ & $114.7^{+1.3}_{-1.4}$ &  $2.18^{+0.20}_{-0.18}$\\
TM8 G2BBPL & 1.34(1446) & $1.36^{+0.05}_{-0.05}$ &  &  & $70.7^{+0.5}_{-0.5}$ & $130.7^{+2.1}_{-2.1}$   & $2.09^{+0.25}_{-0.24}$\\
TM8 G2BBPLe & 1.15 (1444) & $1.98^{+0.09}_{-0.08}$ & $263.8^{+1.7}_{-1.6}$ & $0.404^{+0.027}_{-0.026}$ & $63.4^{+0.7}_{-0.8}$ & $121.2^{+1.5}_{-1.9}$ & $2.27^{+0.23}_{-0.21}$ \\
TM8+XMM G2BBPLe & 1.25 (2002) & $1.96^{+0.07}_{-0.08}$ & $266.1^{+1.6}_{-1.9}$ & $0.406^{+0.015}_{-0.017}$ & $64.3^{+0.5}_{-0.5}$ & $122.3^{+1.3}_{-1.4}$ & $1.86^{+0.19}_{-0.18}$ \\
\hline
\end{tabular}
\tablefoot{Errors are $1\sigma$ confidence levels. The data sets are fitted within $0.2-5$\,keV. The results for the phase-dependent parameters are shown in Fig.\,\ref{f:phrparam}, while the fit residuals for the 2BBPL and Gabs2BBPL fit to TM8 are shown in Fig.\,\ref{f:phrresid}. For the simultaneous TM8+XMM fit a renormalization factor, with respect to the TM8 dataset, of $0.949^{+ 0.005}_{- 0.005}$ for the pn(2015) and of $0.927^{+ 0.005}_{-0.003}$ for the pn(2019) observation was included. 
\tablefoottext{a}{The column density is in units of $10^{20}$\,cm$^{-2}$.}}

\end{table*}

\subsubsection{Phase-resolved spectroscopy \label{s:phrspec}}

The spin-resolved \ero spectra were first fitted using a composite model (2BBPL) built from the combination of two {\tt bbodyrad} components, to model the thermal emission, and a {\tt powerlaw} component to take into account the non-thermal emission at higher energies. A interstellar absorption component was of course also taken into account. Similarly to the phase-averaged analysis, an {\tt edge} and {\tt tbabs} component were multiplied to the spectral model to take into account the feature around $0.26$\,keV, in accordance with the results of Sect.~\ref{s:avspec}. We refer to this model as 2BBPLe, the 'e' indicating the inclusion of the new edge feature.

We set some model parameters to be the same for all phase-binned spectra: the interstellar absorption, the power-law slope, the temperatures of the two blackbody components, and the edge energy. Their best-fit values are listed in Table~\ref{t:phrespar}. In general, the resulting parameter values compare well to the spin phase-averaged fit results (Sect.~\ref{s:avspec}) and the edge and photospheric parameters are not significantly sensitive to the power-law index.

We then performed fits including the Gaussian absorption line and refer to those as G2BBPL and G2BBPLe. All parameters of the Gaussian were allowed to vary freely. All the described models were applied initially to \ero data only and the results for the nonvariable parameters are listed in the first four lines of Table~\ref{f:phrresid}. To improve the accuracy of the model parameters, we also conducted a simultaneous fit with \ero and \xmmn, using also the G2BBPLe model (third line in Table~\ref{f:phrresid} and Fig.~\ref{f:phrparam}). This fit revealed indeed better constrained model parameters and a slightly lower photon index, while there was no significant change of the best-fit values.

Similarly to Fig.~\ref{f:trail}, the residuals of the fits with and without the phase-dependent Gaussian absorption line, as well as their best-fit parameters as given in the first two lines of Table~\ref{t:phrespar}, are shown as an apparent trailed spectrogram in Fig.~\ref{f:phrresid}, only for \ero data. The fit without the absorption line shows strong residuals at energies between $0.5$\,keV and $0.6$\,keV in the phase interval between 0.8 and 1.4. The large $\chi^2_{\rm red}$ value indicates a bad choice of the null hypothesis. The fit with the Gaussian absorption line but without the edge has systematic residuals at 260 eV. The inclusion of both the Gaussian line and the edge reveals a statistically acceptable fit to the data. The two-dimensonal spectrum of the residuals (lowest panel in Fig.~\ref{f:phrresid}) is compatible with pure statistical scatter. 

We show the results regarding the phase-dependent parameters in Fig.\,\ref{f:phrparam}. The variable parameters are the emitting areas and the parameters of the Gaussian absorption line. The radiation radii of the two blackbody components at infinity were computed assuming a distance of $288^{+33}_{-27}$\,pc. The radius values of the cold blackbody component (CBB) peak around phase $0.95$, while the hot blackbodies' (HBB) radius values peak at phase 0.1, indicating a phase shift of $0.1-0.15$ between the cold and hot thermal components. 

As expected and known from the fit to the phase-averaged spectrum, the inclusion of the Gaussian absorption line has some effect on the temperature of the blackbodies and a strong effect on their normalisation, hence emitting radius. The same is true for the edge, which mainlz affects the cold blackbody temperature and hence its radius. Gaining a better understanding of the nature of the features is thus of high relevance for an understanding of the X-ray to ultraviolet SED as discussed by \citet{zharikov+21}. 

Fortunately, the parameters of the Gaussian line are only very weakly affected by the  inclusion or the omission of the edge. Interestingly, there is a dependence of the line parameters with spin phase (see the bottom three panels of Fig.~\ref{f:phrparam}). The absorption is visible for about 60\% of the spin cycle, centered on phase 0.1; it appears redshifted at the beginning and blueshifted at the end of its visibility interval. The same trend is observed in data from TM9 (not shown here), however $30-40$\,eV shifted towards lower energies, due to calibration uncertainties (Sect.~\ref{s:eroccalib}). In general terms, our analysis seems to reveal results that are in overall agreement with the phase-resolved study performed by \citet[][see their Fig.~15]{arumugasamy+18} although the fit strategy was somewhat different. The important difference is that the line parameters are more robustly determined with the new data acquired by us and presented in this work.

\subsubsection{NS atmosphere and condensed surface models \label{s:atmoco}}

The observed X-ray spectrum of \psr can be described with a phenomenological model, but the physical origin of the components that are thought to originate from the stellar surface remain largely unclear. Indeed, X-ray pulsations and the derived blackbody parameters strongly suggest a non-uniform temperature distribution over the NS surface. They do not allow a quantitative interpretation of the observation as such  estimates would be affected by the assumed temperature distribution and the fact that the local spectrum from a magnetized NS surface is known to deviate from a pure blackbody \citep[see, e.g.][]{PL09, VAd05}. In this section we attempt to provide a more physically motivated description of the observed mean spectrum of \psr.  For comparison with the observed phase-averaged spectra we computed models with a range of the angle $\gamma_{\rm B}$ between the magnetic dipole and the rotation axis and use this angle as one of the fit parameters. This is a strong simplification, such a model does not predict pulsations. However, we think that this approach will allow us to find the most promising model for further more sophisticated modeling.

The simplest model applied is a cooling magnetized neutron star with a dipole surface magnetic field covered with a hydrogen atmosphere. We computed a grid of such models, using the method recently employed to fit the thermal spectrum of PSR\,J1957$+$5033 by \citet{Zuzinetal.21}. We assumed the mass and radius to be fixed at $M = 1.4 M_\odot$, $R=12$\,km, the magnetic field strength at the magnetic pole $B_{\rm p} = 10^{13}$\,G. The grid parameter was the bolometric luminosity expressed via the redshifted effective temperature $T^\infty_{\rm eff}$, with $\log T^\infty_{\rm eff}$ from 5.2 to 6.0 with the step 0.1. The neutron star surface was divided into four latitude zones, one including the pole and another the equator. For further details see \citet{Zuzinetal.21}.

The computed grid of theoretical spectra, integrated over the neutron
star surface for $\gamma_{\rm B}=0^\circ, 30^\circ, 60^\circ$, and $90^\circ$, was used to fit the observed phase-averaged 
\ero spectrum (see Sect.~\ref{s:avspec}). The best fit is presented in
Fig.\,\ref{fig:allmod}, and the obtained model parameters are listed in
Table\,\ref{tabl:atm}. The additional spectral components that were
described above, a power law, a Gaussian absorption line and an
absorption edge, were also included, because otherwise the model did not provide a statistically acceptable fit. 

\begin{figure}[t]
\resizebox{\hsize}{!}{\includegraphics{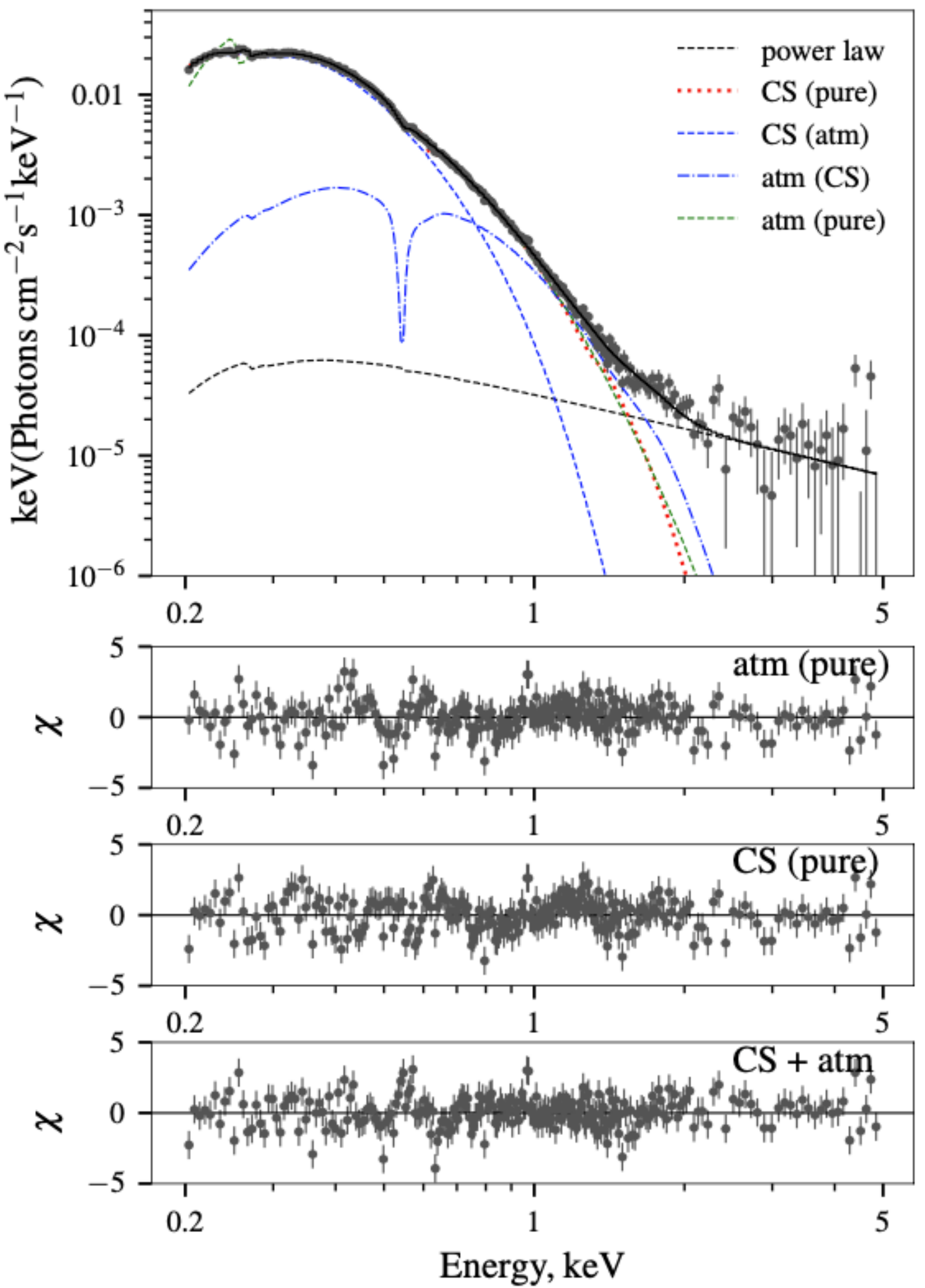}}
\caption{Best fit of the averaged spectrum using three magnetized neutron star model spectra; 1) covered by a hydrogen atmosphere (atm (pure)); 2) covered by an emitting condensed surface (CS (pure)); 3) covered by an emitting condensed surface
with added hot spot covered with a geometrically thin atmosphere (CS + atm; $\Sigma$ = 10 g\,cm$^{-2}$). See the detailed description in the text. The separate contributions of the used components (CS (atm) and atm (CS)) are also shown. The deviations of each model from the observed spectrum are shown separately in the three bottom panels.
\label{fig:allmod}
}
\end{figure}

\begin{table}[t]
\caption{Results of spectral fitting experiments using physically motivated models.
\label{tabl:atm}}
\centering
\begin{tabular}{l|ccc}
\hline\hline
Parameter & Atm & CS & CS + Atm\\
\hline
\rule{0pt}{2.7ex}
  $N_{\rm H}$ (10$^{20}$\,cm$^{-2}$)&3.52$^{+0.1}_{-0.05}$ & 1.04$^{+0.03}_{-0.04}$&1.94$^{+0.04}_{-0.03}$\\
\rule{0pt}{2.7ex}
$D$ (pc)& 59.6$^{+5.9}_{-3.4}$ & 288\tablefootmark{a}& 240$\pm 12$\\
\rule{0pt}{2.ex}
$\Gamma$ &1.98\tablefootmark{a}& 1.98\tablefootmark{a}& 1.98\tablefootmark{a}\\
$K_{\Gamma}$ (10$^{-5}$)& 4.71$^{+0.17}_{-0.2}$& 4.65$^{+0.18}_{-0.21}$& 3.32$\pm    0.17$\\
\rule{0pt}{2.7ex}
$E_{\rm edge}$ (keV)& 0.252$\pm 0.002$& 0.252$^{+0.006}_{-0.010}$& 0.268$\pm 0.003$\\
\rule{0pt}{2.7ex}
$\tau_{\rm edge}$ & 0.61$^{+0.04}_{-0.02}$& 0.07$\pm 0.02$& 0.17$\pm 0.02$\\
\rule{0pt}{2.7ex}
$E_{\rm Line}$ (keV)& 0.589$^{+0.006}_{-0.002}$& 0.537$^{+0.005}_{-0.006}$& -\\
\rule{0pt}{2.7ex}
$\sigma_{\rm Line}$ (keV) & 0.075$\pm 0.005$ & 0.048$^{+0.003}_{-0.004}$ & - \\
\rule{0pt}{2.7ex}
$\tau_{\rm Line}$ &0.038$\pm 0.003$ & 0.044$^{+0.004}_{-0.006}$& -\\
\rule{0pt}{2.7ex}
$T_{\rm p}$ (MK)& 0.355$\pm 0.005$& 1.853$^{+0.01}_{-0.006}$& 1.047$\pm 0.007$\\
\rule{0pt}{2.7ex}
$B_{\rm p}$ (10$^{13}$\,G)& 1\tablefootmark{a} & 25.6$^{+0.3}_{-0.4}$& 10.6\tablefootmark{a}\\
$\gamma_{\rm B}$ &82\degr$^{+0.7\degr}_{-4\degr}$ & 80\degr$\pm$ 2\degr& 90\degr$\pm$ 0\degr\\
$a_{\rm T}$ & 0.25\tablefootmark{a} & 48$^{+2}_{-1}$& 0.25\tablefootmark{a}\\
$T_{\rm sp}$ (MK)& -& -&1.53$\pm 0.01$ \\
$B_{\rm sp}$ (10$^{13}$\,G)& -& -& 10.6\\
$R_{\rm sp}$ (km)& -& -& 0.67$^{+0.06}_{-0.06}$\\
\rule[-1.5ex]{0pt}{4ex}
$\chi^2_{\rm d.o.f.}$ & 1.34 & 1.40 & 1.26\\
\hline 
\end{tabular}
\tablefoot{The values of the temperatures and the magnetic field strengths are given in the neutron star rest-frame, as well as the spot radius $R_{\rm sp}$. The neutron star basic parameters are $M = 1.4 M_\odot$ and $R = 12$\,km for all the models.
\tablefoottext{a}{Fixed parameter.}}
\end{table}

The obtained fit describes the observed spectral shape marginally well, although even with the inclusion of the additional components the quality of the fit is slightly worse than that provided by the phenomenological model. More serious issue is, however, that the derived distance
to the source ($\approx 60$\,pc) is too small compared to the distance obtained from the radio-astrometric parallax, 288$^{+33}_{-27}$\,pc \citep{brisken+03}. Besides the problem with the distance, the model also does not provide self-consistent description for the observed absorption line. The same conclusions were obtained earlier by \citet{arumugasamy+18} using various \xspec spectral models describing the radiation of the magnetized neutron star atmospheres.

To explain the observed absorption feature as proton cyclotron absorption, we considered the possibility that the magnetic field strength could be as high as $B \sim 10^{14}$\,G. Although the characteristic (spin-down) field is an order of magnitude lower, the presence of such strong local fields at the surface cannot be excluded \citep[cf.][]{Tiengo_13,Mereghetti_15}. A plasma envelope of magnetized neutron stars at this high field and at the temperature typical for the surface of \psr ($kT \sim 0.1$\,keV) can be condensed \citep[see, e.g.,][their Fig.~1 and the related discussion]{Tavernaetal.20}. For this reason we considered spectra of magnetized neutron stars with a condensed surface as an alternative physical model to describe the phase-averaged X-ray spectrum. 

The emitting spectrum of the condensed surface can be computed using approaches suggested by  \citet{Turollaetal.04} and \citet{VAd05}, but we  utilize here the analytic approximation for the iron condensed surface \citep{Potekhinetal.12}. A spectrum of the condensed surface is close to blackbody spectrum with two absorption features: one at the electron plasma energy $E_\mathrm{pe}$, and another extending from the ion cyclotron energy $E_{\rm cyc,i}$ to some upper energy $E_{\rm C}\approx E_{\rm cyc,i} + E_\mathrm{pe}^2/E_{\rm cyc,i}$ \citep[see][for details]{VAd05}. The second feature is relatively narrow at high fields, and it can potentially describe the observed absorption line at 0.57 keV. 

However, the integrated spectra of the magnetized neutron stars covered with the condensed surface are close to the single blackbody spectra if a temperature distribution corresponding to the dipole field is assumed. As a result, these models cannot explain the observed spectrum, and  temperature distributions more peaked to the poles have to be considered. This kind of temperature distributions are  possible if we introduce a strong toroidal component of the magnetic field in the crust \citep[see, e.g.][]{PMP06, Suleimanovetal.10}. In these models, most of the heat emerges near the magnetic poles, and the sizes of the polar hot spots depend on the relative contribution of the toroidal component. The temperature distribution is controlled by the related parameter $a_{\rm T}$ according to the formulae given in the above-mentioned papers.  If the toroidal component is negligible, $a_{\rm T} \approx 0.25$, and $a_{\rm T}$ increases with enhancing the toroidal component. 

We developed a simple procedure introduced to \xspec for fitting X-ray spectra with the spectra of magnetized neutron stars with the condensed surface. We fitted the observed phase-averaged \ero spectrum using the same methods and averaging procedure as described above. The best fit has a similar statistical significance as the previous attempt, see Fig.\,\ref{fig:allmod}, and the best fit parameters are also listed in Table \ref{tabl:atm}, in the column labeled `CS'. The fit gives the necessary distance to the pulsar at the reasonable neutron star radius of about 12 km assuming $a_{\rm T} \approx 50$. However, the depression between  $E_{\rm cyc,i}$ and  $E_{\rm C}$ cannot completely describe the observed absorption line at 0.57 keV, and an additional Gaussian absorption line needs to be included to obtain an acceptable fit.

The third model that we were applying to the \ero data includes both, the condensed surface and the model atmosphere. We assume that the whole neutron star surface has a condensed surface, but the regions near magnetic poles with $B_{\rm p} \approx 10^{14}$\,G are covered with a thin hydrogen atmosphere  with the column density $\Sigma = 10$\,g\,cm$^{-2}$. Such an atmosphere is optically thin at continuum photon energies, but it is optically thick near the proton cyclotron line. As a result the emergent spectrum of the thin atmosphere is close to a blackbody with a proton cyclotron line, which can describe the observed absorption feature at 0.57 keV.  This kind of thin atmospheres was suggested for the X-ray dim isolated neutron stars (XDINS, or so-called Magnificent Seven) by \citet{Hoetal.07}, see also \citet{Suleimanovetal.10}.

We computed seven models of thin magnetized hydrogen atmospheres with effective temperatures $T_{\rm eff}$ between 1\,MK and 2.2\,MK with a stepsize of 0.2\,MK and with a normal orientation of the magnetic field of strength $B=10^{14}$\,G. Partial ionization of the atmosphere was taken into account. We used this grid of model spectra to fit the observed spectrum together with the model spectra of the neutron star covered with the condensed surface. The best fit is shown in Fig.\,\ref{fig:allmod}, and the best fit parameters are presented in Table \ref{tabl:atm}, in the third column labeled `CS+Atm'. This fit does not require an additional Gaussian absorption line, but the physical interpretation of the model is not unambiguous. 

The global dipole magnetic field is comparable to the magnetic field strength derived from the observed spin-down, but the magnetic field in the hot spot described by the thin atmosphere is ten times larger. This may be explained by a complex structure of the field near the surface, with higher multipole components being much stronger than the dipole component. Such models are widely discussed in view of mounting body of observational and theoretical evidence that a complex field topology should be the rule rather than the exception \citep[see, e.g.,][and references therein]{Viganoetal.21}.  The thin hydrogen atmosphere can be the result of the  nuclear spallation processes, which may present a self-regulating mechanism for producing a thin hydrogen atmosphere above a condensed iron surface (as suggested by \citealt{Hoetal.07} in the case of RX J1856.5$-$3754). We note, that the multi-component field model with B$\approx 10^{14}$\,G was also discussed  by \citet{arumugasamy+18}.

\section{Discussion and conclusion \label{s:disc}}

We have presented an in-depth analysis of a long, 100 ks, uninterrupted \ero observation of the nearby radio pulsar \psr. This was the first \ero observation after the (longer than originally planned) comissioning phase performed with seven TMs of a stellar target. For cross-calibration purposes, the \ero observations were accompanied by simultaneous observations with \xmmn. Our analysis further benefits from the inclusion of archival \nic and \nus data, the former for timing, the latter to better constrain the non-thermal emission. TM1 did not deliver science-grade data; for the spectral analysis, we only consider the detectors not affected by the optical leak (TM8).

We have characterised the capabilities (and limitations) of \ero for timing studies of fast spinning neutron stars such as \psr (spin period of 385\,ms). Being performed at such an early phase of the mission, it was not possible to obtain an absolute time calibration of our observation. The relative timing accuracy was found to be of order $5 \times 10^{-7}$\,s. Phase-folded lightcurves were eventually created using the \ero-determined period in various energy bands and found to be fully consistent with those from \xmmn, despite the much lower time resolution of \ero.

While addressing the question about the phase offset between the X-ray maximum and the radio pulse, we encountered unexpected timing uncertainties in current radio ephemerides, despite their very high precision. We then established a new X-ray ephemeris based on two \xmmn and several \nic observations, which covers the time interval from 2015 to 2020 with sufficient accuracy. Furthermore, we noticed a strong evolution of the X-ray pulse TOAs beyond the 2015 epoch, which could be related to one or several glitches that have, up to this point, remained unnoticed. Our target is a well-known glitching pulsar \citep{espinoza+11}, therefore this interpretation is perhaps not unlikely. We did not establish, however, the X-ray-to-radio phase relation, awaiting further improvements of the ephemeris. On this regard, it will be extremely beneficial to have the involvement of the radio pulsar community in a joint analysis, for which we provide in this work the necessary input data from the X-ray side. 

The main motivation to select \psr in the performance and verification phase of \ero was the tentative identification of an absorption feature at around 540\,eV in a long \xmmn observation reported by \cite{arumugasamy+18}. The new observations add further knowledge about this feature. We firstly characterised the phase-averaged spectrum. The phenomenological model, which was applied here and follows the description established by \citet[e.g.~][]{arumugasamy+18}, consists of two blackbody component, a hard power-law tail and a superposed Gaussian at soft X-rays (G2BBPL). The Gaussian is highly significant and can be regarded as securly etablished. The parameters of the feature, its energy and width, are revised and their accuracy improved. 

Our modeling revealed an additional feature at soft energies, here described as an edge at about $260-265$\,eV, which may not be instrumental (model G2BBPLe). \cite{zharikov+21} described a feature at $\sim$0.3 keV when extending the analysis of \xmmn data to energies below 0.3 keV. While their fit formally clearly was improved through the inclusion of this additional line, they are cautious about its existence and seek for an independent detection with \ero. This paper presents the observational status of this feature for this particular star and we are still cautious. In this regard, the analysis of additional observations of bright isolated neutron stars, namely the Magnificent Seven, performed by \ero at several occasions since launch, will hopefully settle the issue. If confirmed to be astronomical, not instrumental, it will reveal sought-for boundary conditions for future theoretical modeling of isolated neutron stars. 

We have presented three different types of model spectra and applied those to the multi-mission spectral data: a magnetised atmosphere, a condensed surface, and a mixed model. All three describe the shape of the continuum spectrum well but only the last one provides some natural explanation for the occurrence of an absorption line.

The first model tested was a semi-infinite atmospheres with temperature distributions corresponding to a dipole magnetic field. The model spectra resemble the observed one at relatively low effective temperatures implying a short distance to the star. Should the X-ray emitting region be smaller than the whole stellar surface, the implied distance would even be smaller. The inapplicability of standard \xspec neutron star atmosphere models was noticed also by \citet{arumugasamy+18}.

If the magnetic field is much higher than 10$^{13}$\, G, other models are suitable and have been tested: a condensed surface and a geometrically thin atmosphere. The local spectrum of the condensed surface is close to blackbody spectrum. Therefore, we need a special temperature distribution which mimics two blackbodies -- a bright pole with a relatively cold rest surface -- to obtain a reasonable spectral fit. 

In the third model, the whole neutron star surface is covered by the condensed surface, and its radiation mimics that of the cold blackbody component in the phenomenological model. We then introduced the bright spot covered with the thin atmosphere. This second component mimics the hot blackbody component in the two-blackbody fit. In addition it contains the proton cyclotron line that explains the observed absorption feature, whose existence was definitely confirmed in this study \citep[see the discussion of their tentatively identified feature by][]{arumugasamy+18} .

In both cases of a pure condensed surface and the condensed surface plus thin atmosphere, the surface effective temperatures are significantly higher than for the semi-infinite atmospheres. Therefore, the observed X-ray flux level can be provided by the neutron star at the pulsar's distance of 288\,pc. Our analysis of the phase-averaged spectrum is intended to test different physical models and possibly pre-determine some of the spectral parameters. Construction of detailed, physically motivated fits to the phase-resolved spectra deserves a separate study, but it is beyond the scope of the present paper.

The large number of photons that could be collected with \ero allowed to study the phase dependent behavior of the main spectral parameters. It could be traced through 60\% of the spin cycle. This observation rules out an instrumental origin of the feature. The line is centered on phase 0.1, and shows $1\sigma$ variability of all its parameters: its location, width, and derived optical depth. Similarly, the behaviour of the blackbody components could be characterised through our phase-resolved spectroscopic study. The cold component reaches its maximum at phase $\sim$0.9 whereas the hot component peaks at phase $\sim$0.05. This behaviour is clearly different from that described by \citet{deluca+05}, who find an anti-correlation between the radii of the cold and the hot blackbody components. The higher statistics obtained with the deep \xmmn and \ero observations clearly establish a more complex picture. The absolute value of the blackbody radii (not their phase relation) is strongly dependent on the inclusion of the absorption line at 570\,eV in the modeling. The nature of this absorption line remains, however, uncertain. Difficulties to achieve a unique interpretation arise from complex temperature distributions over the stellar surface and corresponding kinks in the continuum spectra \citep[e.g.][]{vigano+14}, from calibration uncertainties, from uncertainties of the correct spectral model and the limited energy coverage of the spectra. 

Despite those difficulties a few important constraints can be derived regarding the location where this line originates. Is it far (magnetospheric) or near (atmospheric)? The line is seen only for $\sim$60\% which is perhaps simpler to understand if it would be near ($h << R_{\rm NS}$). However, \citet{Tiengo_13} discuss a possible far location ($R\sim 3R_{\rm NS}$) of the absorber in the magnetar SGR 0418$+$5729 to explain the phase-variable absorption feature. If applied to our target, the implied magnetic field strength from the proton cyclotron interpretation would not inform about the field at the surface, which would be significantly higher and in the magnetar regime. Observationally, the line visibility is centered on phase 0.1, i.e.~it coincides in phase with the maximum of the hot blackbody emitting region. One could thus speculate about a direct relation between the two. This on the other hand is pre-mature. The hot blackbody is seen at all phases, its emitting radius varies by $\sim$30\% only, whereas the absorption line is clearly constrained in phase. In this regard, it is instructive to check Figs.~3 and 4 of \cite{beloborodov02} who has sketched visibility periods and lightcurve shapes of blackbody pulse profiles as a function of the inclination $i$ of the spin axis and the spot latitude $\theta$. The lightcurve of the hot blackbody resembles their class I, i.e.~a combination of not too high values of both angles. The `lightcurve' (here visibility curve) of the absorption feature resembles their class II, which would imply a larger spot latitude than the blackbody for a given inclination. Hence the hot blackbody and the absorption like are likely spatially disjunct despite the similarities in their phase-dependence.

Phase-resolved lightcurves for photon energies below $\sim$2\,keV could be established with very high signal-to-noise ratio. At energies below 0.7\,keV, they are highly asymmetric and might be indicative of a third spot or a more complex temperature distribution over the surface of the star. One may attempt to model this using MEM-inversion techniques similar as is done to map spots on active stars \citep[e.g.][]{berdyugina+02}, once a better understanding of the continuum emission processes are achieved. These studies will also benefit from the larger body of \ero observations of XDINS (Pires et al., in preparation).

The evidence for narrow absorption features at similar energies have been reported in the RGS spectra of several thermally emitting isolated neutron stars \citep[see][and references therein]{hambaryan+09, hohle+12}, however their interpretation remains unclear. In particular, the analysis of \citet{pires+19} of the nearby XDINS RX~J1605.3+3249, based on a recent \xmmn large programme and on archival RGS observations of the source extending back to 2002, revealed that evidence for the line is only found in the early observations. For \psr we found weak evidence for a line in the RGS spectra alone, its energy and parameters being however consistent with those found in \ero and pn data. 

\begin{acknowledgements}
Support of this work by the German DLR under contracts 50 QR 1604, 50 QR 2104 and 50 OX 1901 is gratefully acknowledged. AMP gratefully acknowledges support from the Chinese Academy of Science's President International Fellowship Initiative (CAS PIFI 2019VMC0008). VS thanks the Deutsche Forschungsgemeinschaft (DFG) for financial support (grant WE 1312/53-1). His work has also been supported by the grant 14.W03.31.0021 of the Ministry of Science and Higher Education of the Russian Federation. The work of AYP was partially supported by the Russian Foundation for Basic Research (RFBR) according to the research project 19-52-12013.

We thank the \xmmn\ project scientist, Dr.~Norbert Schartel, for the generous allocation of observation time and the ESAC-SOC for simultaneous scheduling of the observations.

We thank Chris Flynn, Matthew Bailes and Marcus Lower from the Swinburne University of Technology for communicating their latest ephemeris for B0656+14 prior to publication.

This work is based on data from \ero, the primary instrument aboard SRG, a joint Russian-German science mission supported by the Russian Space Agency (Roskosmos), in the interests of the Russian Academy of Sciences represented by its Space Research Institute (IKI), and the Deutsches Zentrum für Luft- und Raumfahrt (DLR). The SRG spacecraft was built by Lavochkin Association (NPOL) and its subcontractors, and is operated by NPOL with support from the Max Planck Institute for Extraterrestrial Physics (MPE).

The development and construction of the \ero X-ray instrument was led by MPE, with contributions from the Dr. Karl Remeis Observatory Bamberg \& ECAP (FAU Erlangen-Nuernberg), the University of Hamburg Observatory, the Leibniz Institute for Astrophysics Potsdam (AIP), and the Institute for Astronomy and Astrophysics of the University of Tübingen, with the support of DLR and the Max Planck Society. The Argelander Institute for Astronomy of the University of Bonn and the Ludwig Maximilians Universität Munich also participated in the science preparation for ero.

The \ero data shown here were processed using the eSASS/NRTA software system developed by the German \ero consortium.

\end{acknowledgements}

\bibliography{psr}


\begin{appendix}

\section{TM1 malfunction \label{s:tm1}}

The anomalous behaviour of TM1 was noticed early and could not be fixed during the various pipeline processing stages of this observation, which took into account calibration updates and ever improved software versions. In the most recent processing (version c001) which enters the public area, the overall number of photons in TM1 was only about 60\% of that in the other cameras, a coarsely binned lightcurve (50\,s bin size) in original time sequence was much noisier than for the other TMs and showed a low-frequency modulation of the mean rate, and a GTI-filtered lightcurve contained no photons at all. The reason for the malfunction of TM1 during this observation is not known, we exclude the data from TM1 from the scientific analysis presented in this paper.

\section{The time system of \srg and \ero \label{s:timeshift}}

NPOL/Lavochkin in cooperation with Roscosmos operates the SRG spacecraft and controls the central spacecraft clock via radar signals. Such measurements are performed routinely since November 3, 2019, hence almost three weeks after the observations of B0656, with the simple consequence that an absolute value for the photon arrival times of the pulsar as observed with \srgero cannot be given. Time shifts are applied to the central spacecraft clock due to some instability of the central quartz. The size of it and the possible effect on our observations are described here.

Once the offsets between UTC and OBT (onboard time) reach a limit of 1 sec, time shifts are applied to keep the offsets always between 0 and 1 s. The history of such shifts in the time interval between launch and end of year 2020 is documented in Table~\ref{t:off}. The offset history cleaned for the time shifts is shown in Fig.~\ref{f:tdel}. For data representation and polynomial fits applied to the data the time axis was re-normalized to 0 at MJD 59000. Time offset errors were assumed to be 0.002 s before MJD 59100, and 0.0005s after that date. Those values were estimated from the scatter of the daily measured clock drift by us. A linear fit to the measured drift (upper graph in Fig.~\ref{f:tdel}) does not provide an acceptable representation of the data. Only a cubic fit applied to the data leaves mainly scatter in the $O-C$ residuals (shown in the lower panels of the two graphs). In the context of the current paper only the linear term is relevant, which is $0.0119660(5)$ s d$^{-1}$, and $ 0.011933(1)$ s d$^{-1}$, respectively, for the two fits shown.

\begin{figure}[h]

\resizebox{\hsize}{!}{\includegraphics{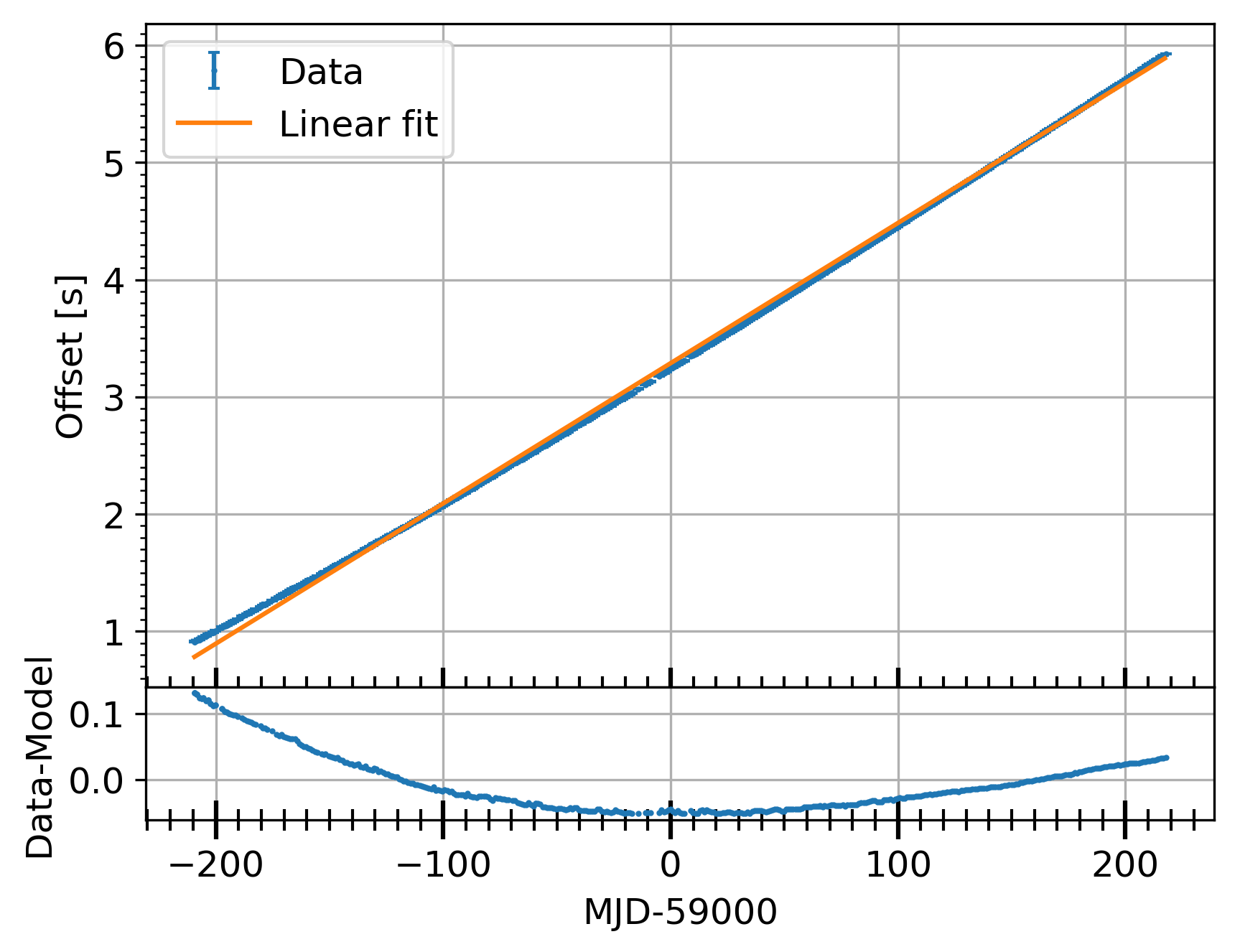}}

\resizebox{\hsize}{!}{\includegraphics{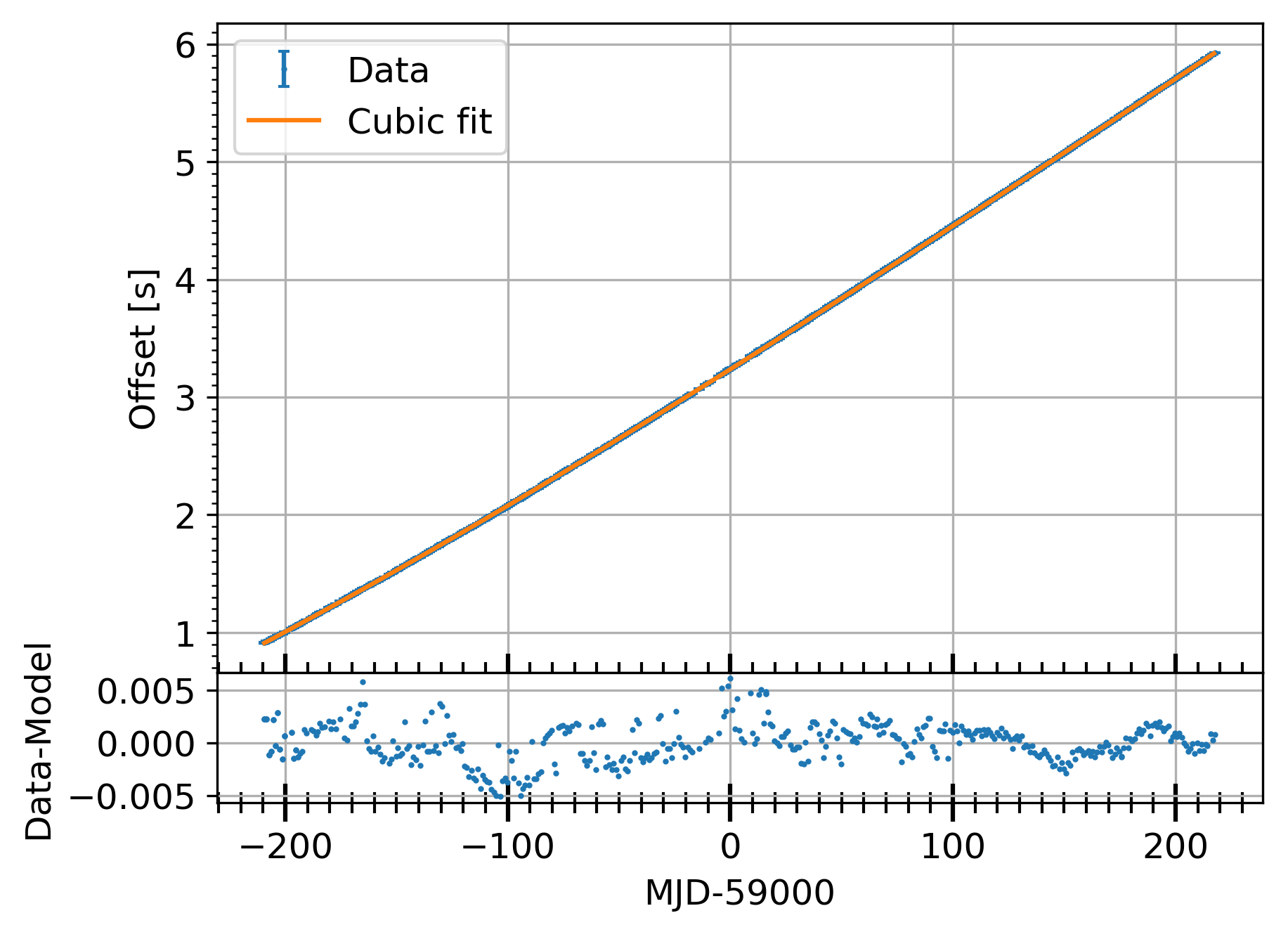}}

\caption{Linear and cubic fits to the accumulated offsets between UTC and the spacecraft clock. \label{f:tdel}
}
\end{figure}

\begin{table}
\caption{Time-shift history of the SRG quartz
\label{t:off}}
\centering 
\begin{tabular}{lrrr}
\hline\hline
 & \multicolumn{2}{c}{Onboard execution time} & Value \\
\cline{2-3}
 & \multicolumn{2}{c}{(UTC)} & (s) \\
\hline
1 & 2019-07-19 & 16:57:16.4 & 0.094 \\
2 & 2019-07-19 & 16:57:34.8 & -2.000\\
3 & 2019-10-14 & 23:00:45.0 & 0.345\\
4 & 2019-11-03 & 04:55:00.0 & 0.900\\
5 & 2020-01-29 & 14:39:23.7 & 0.925\\
6 & 2020-04-01 & 17:20:00.0 & 0.625\\
7 & 2020-06-15 & 22:52:20.3 & 0.960\\
8 & 2020-08-04 & 17:02:22.5 & 0.620\\
9 & 2020-10-05 & 21:10:00.0 & 0.625\\
10 & 2020-11-22 & 18:43:34.5 & 0.740\\
\hline
\end{tabular}
\tablefoot{We list the shifts' onboard execution times in UTC and magnitude in seconds. The correction in entry 3 was applied during the observation of \psr}
\end{table}

\section{X-ray ephemeris based on \nic and \xmmn data}

In the table \ref{t:ephlit} we list some of the relevant parameters extracted from the parameter files describing the available timing solutions. The values from Lower et al.~were made kindly available to us prior to publication.

\begin{table*}
\caption{Ephemeris parameter files (abridged) used in this paper\label{t:ephlit}}
\begin{tabular}{llll}
Parameter & Ray+11 & JK18 & Lower+20\\
\hline
F0        & $         2.59794980081  \pm 6.4\mbox{e-10}$ &
                     $2.597948806669 \pm 1.1\mbox{e-11}$ & 
                     $2.597884228896 \pm 2.9\mbox{e-11}$ \\
F1        &          $-3.709345\mbox{e-13}  \pm 5.8\mbox{e-18}$ & 
                     $-3.7093504\mbox{e-13} \pm 1.7\mbox{e-19}$ & 
                     $-3.7078343\mbox{e-13} \pm 5.0\mbox{e-19}$\\
PEPOCH    &           55555                     & 55586 & 57600 \\
START     &           54689.8898927             & 54505.459128 & unav.\\
FINISH    &           56577.7417393             & 56670.499208 & unav.\\
TZRMJD &  55676.655274553879281  & 55575.484087148401098 & 58129.56652957678185\\
UNITS     &           TDB                       & TDB & TCB \\
TIMEEPH   &           FB90                      & FB90 & IF99 \\
EPHEM     &           DE405                     & DE405 & DE430 \\
NTOA      &           68                        & 70 & 245 \\
\hline
\end{tabular}
\end{table*}

In the table ~\ref{t:xraytoa} below we list the  \nic (abbreviation NI) and \xmmn (abbreviation X) observations (identified with the abbreviations NI or X followed by the respective observation ID) and the derived pulse arrival time that were used for the determination of the X-ray ephemeris of Section~\ref{s:timingsolution}

\begin{table*}
\small
\caption{Summary of observations and times-of-arrival (TOA) for pulsar ephemeris. All times are given in TDB.
\label{t:xraytoa}}
\centering
\begin{tabular}{c|p{100mm}|r|r}
\hline\hline
Group & \multicolumn{1}{c|}{ObsID} & \multicolumn{1}{c|}{TOA} & Covered Interval \\
 & & \multicolumn{1}{c|}{(MJD)} & \multicolumn{1}{c}{(MJD)} \\
\hline
1 & X0762890101 & 57284.84532588(3) & 57284.85-57286.33\\
2 & NI1020130102 & 58040.05190366(7) & 58040.05-58040.32\\
3 & NI1020130103, NI1020130116, NI2020130101, NI2020130114, NI1020130109, NI2020130103, NI2020130116, NI1020130101, NI1020130114, NI2020130119 & 58039.92321915(2) & 58039.92-58843.90\\
4 & NI1020130104 & 58042.3048977(1) & 58042.30-58042.83\\
5 & NI1020130106, NI2020130105 & 58044.03145455(3) & 58044.03-58787.81\\
6 & NI1020130108 & 58046.02583683(4) & 58046.03-58046.76\\
7 & NI1020130113 & 58061.0413178(1) & 58061.04-58061.36\\
8 & NI1020130115 & 58063.02898660(7) & 58063.03-58063.81\\
9 & NI1020130118, NI1020130119, NI1020130111 & 58050.99369333(2) & 58050.99-58091.35\\
10 & NI1020130122, NI2020130135 & 58111.17533634(7) & 58111.18-58868.80\\
11 & NI1020130123, NI1020130136, NI3020130103, NI2020130121, NI2020130134, NI2020130129, NI1020130117 & 58089.53825834(3) & 58089.54-58917.81\\
12 & NI1020130128 & 58140.25374991(7) & 58140.25-58140.71\\
13 & NI1020130133, NI1020130126, NI2020130138, NI2020130125 & 58115.55081734(5) & 58115.55-58892.76\\
14 & NI1020130138, NI1020130125, NI1020130130, NI1020130141 & 58114.01299548(4) & 58114.01-58196.94\\
15 & NI1020130140 & 58195.02269920(4) & 58195.02-58195.98\\
16 & X0853000201 & 58770.54109986(3) & 58770.54-58771.38\\
17 & NI2020130107, NI2020130112 & 58789.29194947(5) & 58789.29-58802.01\\
18 & NI2020130108, NI1020130105, NI1020130110 & 58043.00250000(2) & 58043.00-58790.97\\
19 & NI2020130109 & 58791.02844419(9) & 58791.03-58791.30\\
20 & NI2020130110, NI1020130107, NI1020130112, NI2020130118, NI2020130117 & 58045.07212636(2) & 58045.07-58823.89\\
21 & NI2020130111, NI2020130104 & 58783.04257321(6) & 58783.04-58793.41\\
22 & NI2020130113, NI2020130106 & 58788.38572341(7) & 58788.39-58805.56\\
23 & NI2020130115 & 58813.3644498(1) & 58813.36-58813.37\\
24 & NI2020130120, NI3020130102, NI1020130146 & 58243.09235924(5) & 58243.09-58915.29\\
25 & NI2020130123, NI2020130136, NI3020130101, NI1020130121, NI1020130134, NI1020130129, NI3020130109, NI3020130106 & 58110.13366121(3) & 58110.13-58948.48\\
26 & NI2020130126 & 58859.69284804(8) & 58859.69-58859.97\\
27 & NI2020130130, NI3020130107, NI1020130143, NI1020130127 & 58116.11639425(4) & 58116.12-58922.71\\
28 & NI2020130131, NI2020130124, NI1020130142 & 58197.01438528(5) & 58197.01-58864.92\\
29 & NI2020130133 & 58866.03678855(9) & 58866.04-58866.63\\
30 & NI3020130104, NI1020130139 & 58194.11049189(3) & 58194.11-58919.88\\
31 & NI3020130105, NI2020130127, NI2020130132, NI1020130131, NI1020130124 & 58113.35834534(2) & 58113.36-58920.85\\
32 & NI3020130108, NI2020130137, NI2020130122, NI1020130135, NI1020130120, NI2020130128, NI1020130137 & 58093.78887617(4) & 58093.79-58927.75\\
\hline
\end{tabular}
\end{table*}

\begin{table*}
\small
\caption{Mean best-fit parameters and percentage deviation according to \ero detector type and pattern
\label{t:mdev_edge2bbpgabs}}
\centering
\begin{tabular}{l|ccc|ccc|ccc|ccc}
\hline\hline
Parameter & \multicolumn{3}{c|}{Mean value (TM 2 3 4 6)} & \multicolumn{3}{c|}{Mean value (TM 5 7)} & \multicolumn{3}{c|}{\% (TM 2 3 4 6)} & \multicolumn{3}{c}{\% (TM 5 7)} \\ 
\cline{2-4}\cline{5-7}\cline{8-10}\cline{11-13}
 & s & d & sdtq & s & d & sdtq & s & d & sdtq & s & d & sdtq\\ 
\hline
$\nh$\tablefootmark{a} & $2.13(27)$ & $1.8(3)$ & $1.96(25)$ & 1.80(19) & $1.7(4)$ & $1.60(18)$ & 30 & 50 & 30 & 17 & 40 & 18\\
$\epsilon$ (eV) & $265.2\pm2.0$ & $255\pm14$ & $263\pm5$ & $218.5\pm2.2$ & $221.5\pm1.6$ & $218.0\pm2.5$ & 15 & 13 & 15 & 2.9 & 17 & 8\\
$\tau$ & $0.43(8)$ & $0.51(9)$ & 0.39(8) & 0.98(13) & 1.64(5) & 1.01(10) & 12 & 9 & 12 & 7 & <1 & 5\\
$kT_1$ (eV) & $60\pm3$ & $63\pm4$ & $63\pm4$ & $60.3\pm1.1$ & $49\pm5$ & $54.1\pm2.7$ & 2.1 & 16 & 5 & 1.6 & 1.1 & 1.8\\
$R_1$ (km) & $15.4\pm2.7$ &  $13.6\pm2.6$ & $13.8\pm2.6$ & $19.5\pm2.2$ & $19\pm7$ & $19\pm3$ & 50 & 40 & 50 & 20 & 5 & 15\\
$kT_2$ (eV) & $118\pm5$ & $122\pm5$ & $121\pm5$ & $119\pm5$ & 129.5(3) & $116\pm3$ & 50 & 40 & 50 & 18 & 60 & 26\\
$R_2$ (km) & $1.32(24)$ & $1.16(21)$ & $1.24(22)$ & 1.45(22) & 0.65(3) & 1.30(19) & 50 & 40 & 40 & 24 & 8 & 23\\
$\Gamma$ & $1.96(17)$ &   $2.74(15)$ & 2.44(19) & 1.80(13) & 2.35(9) & 2.45(16) & 23 & 15 & 20 & 11 & 6 & 10\\
$\epsilon$  & $566\pm14$ & $565\pm11$ & $569\pm10$ & $485\pm22$ & $600\pm40$ & $530.0\pm2.5$ & 7 & 6 & 5 & 7 & 10 & <1\\
$f_{\rm X}$\tablefootmark{b}& $1.24(9)$ & $1.17(3)$ & 1.11(3) & 1.08(4) & 0.290(10) & 0.94(10) & 20 & 8 & 6 & 6 & 5 & <1\\
\hline 
\end{tabular}
\tablefoot{The model fit to the data is {\tt tbabs((bbodyrad+bbodyrad+pow)gabs)edge}. The reduced chi-square values of the individual fits range within 1 and 1.5 for $132-171$ degrees-of-freedom for detectors 2, 3, 4 and 6 and within $1.3-2.1$ for $128-169$ for detectors 5 and 7.
\tablefoottext{a}{The column density is in units of $10^{20}$\,cm$^{-2}$.}
\tablefoottext{b}{The observed model flux is in units of $10^{-11}$\,erg\,s$^{-1}$\,cm$^{-2}$ in energy band 0.2--12 keV.}}
\end{table*}

\section{Cross-calibration of the \ero detectors \label{s:eroccalib}}

For cross-calibration purposes, we perform simple fits of the phenomenological model discussed in Sect.~\ref{s:avspec} to the spectra of \psr in each of the six TMs. The same analysis and fitting procedure as described in Sect.~\ref{s:avspec} is applied here. In Table~\ref{t:mdev_edge2bbpgabs} we list the mean results over \ero detector type (with and without on-chip filter) and for various pattern combinations (s, d, sdtq). The average percentage deviation from the mean values is also listed for the two detector groups. For a given pattern, these are typically smaller for TMs 5 and 7, while larger dispersion of the best-fit parameters is consistently observed for doubles in both groups. The worse fit quality (larger chi-square values) are obtained for the detectors affected by the optical light leak (TMs 5 and 7). The largest inconsistencies between patterns are observed for the parameters of the Gaussian absorption and the overall model flux of detectors 5 and 7.

In Table~\ref{t:crosscalib} we compare the results of a single ``stacked'' spectra, extracted from the concatenated TM8 and TM9 event files, with those where, according to detector type, 5 and 2 individual detector data sets are fitted simultaneously in XSPEC. These results are dubbed TM2346 and TM57 in Table~\ref{t:crosscalib}; no filtering in photon pattern is done. Similarly, we show the results of TM0 against TM234567 and TM89 (6 and 2 data sets, respectively). Whenever the data sets were fitted simultaneously we allowed for a re-normalisation factor in XSPEC to account for calibration uncertainties between the instruments. The strength of the edge component in each data set was also allowed to vary independently of each other. As usual, the source and background spectra of the stacked data sets, as well as the ancillary and response files, were produced with the eSASS task {\it srctool}. As a reference for the best-fit parameters (however without the edge component), the fit results of the simultaneous fit of the two EPIC-pn data sets described in Sects.~\ref{s:xmmreduction} and \ref{s:avspec} are also shown in Table~\ref{t:crosscalib}.

Clearly, the two types of \ero instrument converge to different fit solutions. In particular, the best-fit blackbody temperatures of the TM8 data sets agree with pn within $1.1-6$\%, whereas it is $10-16$\% lower for the data sets without the on-chip filter. This is likely due to the altered energy scale of these two detectors as a result of the optical leak (Sect.~\ref{s:obs}). The parameters of the edge component are also remarkably inconsistent, the edge being detected at a 20\% lower energy and at much higher optical depth in TM9 than in TM8. This argues against stacking the two detectors types into a single data set, as it may bias the parameter estimation (that is, the fit does not necessarily converge to an average solution). On the other hand, the analysis shows that the adoption of TM8 or TM9 instead of the corresponding data sets 2, 3, 4, and 6 and 5 and 7, produce results that are completely consistent within the errors. Moreover, the fit results of Table~\ref{t:crosscalib} agree well with the mean parameters of Table~\ref{t:mdev_edge2bbpgabs}, as it should be expected. 

\begin{table*}[!b]
\small
\caption{Cross-calibration of the two different types of \ero detectors
\label{t:crosscalib}}
\centering
\begin{tabular}{clrrrrrrrrrrc}
\hline\hline
Instrum. & $\chi^2_\nu$\,(dof) & \multicolumn{1}{c}{$\nh$\tablefootmark{$a$}} & \multicolumn{2}{c}{Edge} & \multicolumn{1}{c}{$kT_1$} & \multicolumn{1}{c}{$R_1$\tablefootmark{b}} & \multicolumn{1}{c}{$kT_2$} & \multicolumn{1}{c}{$R_2$\tablefootmark{b}} & \multicolumn{1}{c}{$\Gamma$} & \multicolumn{1}{c}{$\epsilon$} & \multicolumn{1}{c}{$\sigma$} & \multicolumn{1}{c}{$f_{X}$\tablefootmark{c}} \\
\cline{4-5}
 & & & ($\epsilon$; eV) & \multicolumn{1}{c}{$\tau$} & \multicolumn{1}{c}{(eV)} & \multicolumn{1}{c}{(km)} & \multicolumn{1}{c}{(eV)} & \multicolumn{1}{c}{(km)} & & \multicolumn{1}{c}{(keV)} & \multicolumn{1}{c}{(eV)} & \\
\hline
pn & 1.2\,(110) & $ 4.3_{-0.9}^{+1.2}$ & & & $ 65.5_{-2.9}^{+2.5}$ & $ 14_{-8}^{+13}$ & $ 128_{-4}^{+5}$ & $ 0.8_{-0.4}^{+0.5}$ & $ 1.85_{-0.12}^{+0.12}$ & $ 539_{-26}^{+18}$ & $ 124_{-23}^{+26}$ & 0.97\\ 
\hline
$8$ & 1.1\,(243) & $1.68_{-0.15}^{+0.16}$ & $260.0_{-2.0}^{+2.0}$ & $0.32_{-0.03}^{+0.04}$ & $64.8_{-2.0}^{+2.1}$ & $11_{-4}^{+5}$ & $120_{-3}^{+3}$ & $1.1_{-0.5}^{+0.5}$ & $2.50_{-0.20}^{+0.3}$ & $571_{-4}^{+4}$ & $67_{-7}^{+8}$ & 1.11\\

$9$ & 1.7\,(202) & $1.54_{-0.12}^{+0.12}$ & $218.0_{-2.0}^{+2.0}$ & $0.98_{-0.06}^{+0.06}$ & $54.5_{-1.0}^{+1.0}$ & $17_{-6}^{+7}$ & $115.0_{-2.0}^{+2.0}$ & $1.3_{-0.4}^{+0.5}$ & $2.4_{-0.3}^{+0.3}$ & $527_{-7}^{+7}$ & $24_{-10}^{+9}$ & 0.93\\ 

$0$ & 1.3\,(270) & $1.23_{-0.10}^{+0.10}$ & $225.0_{-2.0}^{+2.0}$ & $0.53_{-0.04}^{+0.04}$ & $63.9_{-1.3}^{+1.4}$ & $10_{-4}^{+4}$ & $121.0_{-2.0}^{+2.0}$ & $1.1_{-0.4}^{+0.4}$ & $2.30_{-0.20}^{+0.20}$ & $566_{-4}^{+4}$ & $62_{-6}^{+6}$ & 1.03\\ 

$2346$ & 1.3\,(711) & $1.64_{-0.15}^{+0.15}$ & $258.0_{-2.5}^{+2.4}$ & $0.41_{-0.04}^{+0.04}$ & $65.0_{-1.9}^{+2.0}$ & $11_{-4}^{+5}$ & $120_{-3}^{+3}$ & $1.1_{-0.5}^{+0.5}$ & $2.44_{-0.24}^{+0.25}$ & $572_{-4}^{+4}$ & $66_{-7}^{+7}$ & 1.09\\ 

$57$ & 2.2\,(348) & $1.49_{-0.12}^{+0.12}$ & $216.7_{-1.8}^{+2.0}$ & $1.02_{-0.06}^{+0.06}$ & $54.7_{-1.0}^{+1.1}$ & $17_{-6}^{+6}$ & $115.2_{-2.1}^{+2.1}$ & $1.3_{-0.4}^{+0.5}$ & $2.45_{-0.28}^{+0.3}$ & $528_{-7}^{+7}$ & $25_{-9}^{+9}$ & 0.93\\ 

$234567$ & 2.0\,(1069) & $1.15_{-0.09}^{+0.09}$ & $222.7_{-1.9}^{+2.0}$ & $0.85_{-0.05}^{+0.05}$ & $62.9_{-1.1}^{+1.3}$ & $12_{-4}^{+4}$ & $119.7_{-1.8}^{+2.0}$ & $1.2_{-0.4}^{+0.4}$ & $2.39_{-0.18}^{+0.19}$ & $565_{-4}^{+4}$ & $60_{-6}^{+6}$ & 1.10\\ 

$89$ & 2.4\,(455) & $1.27_{-0.13}^{+0.09}$ & $226.3_{-3}^{+2.1}$ & $0.67_{-0.04}^{+0.05}$ & $62.0_{-1.1}^{+1.2}$ & $12_{-4}^{+4}$ & $118.8_{-1.7}^{+1.8}$ & $1.2_{-0.4}^{+0.4}$ & $2.42_{-0.18}^{+0.18}$ & $564_{-4}^{+4}$ & $58_{-6}^{+6}$ & 1.03\\ 

\hline
\end{tabular}
\tablefoot{All valid patterns in the energy range $0.2-5$\,keV are considered. 
\tablefoottext{a}{The column density is in units of $10^{20}$\,cm$^{-2}$.}
\tablefoottext{b}{The radiation radius at infinity of each blackbody component is computed assuming a distance of 288\,pc.}
\tablefoottext{c}{The observed model flux is in units of $10^{-11}$\,erg\,s$^{-1}$\,cm$^{-2}$ in energy band 0.2--12 keV.}}
\end{table*}

\end{appendix}

\end{document}